\documentclass[a4paper,12pt]{article}
\pdfoutput=1
\usepackage{graphicx,subfigure,amsmath,amssymb,cases}
\usepackage{cite}
\usepackage{color}

\newlength{\dinwidth}
\newlength{\dinmargin}
 \def \kb{{\bf k}_{\bot}}
   \def\b{\beta}

\allowdisplaybreaks
\begin{document}
\title{\bf Study of $ B_{u,d,s}  \to K_0^*(1430) P $ and $K_0^*(1430) V$  decays  within QCD factorization }
\author{Lili Chen$^{a}$, Mengfei Zhao$^{a}$, Yunyun Zhang$^{a,b}$ and Qin Chang$^{a}$\\
{ $^a$\small Institute of Particle and Nuclear Physics, Henan Normal University, Henan 453007, China}\\
$^b$\small Nanyang Institute of Technology, Henan 473004, China
}
\date{}

\maketitle

\begin{abstract}
We study the nonleptonic charmless $B_{u,d,s}  \to K_0^*(1430)P$ $(P=K\,, \pi)$ and $ K_0^*(1430)V$ $ (V=K^*\,, \rho\,, \omega\,, \phi)$ decays. The amplitudes are calculated within the QCD factorization, and the non-perturbative quantities are evaluated by using a covariant light-front approach.   The branching fractions and CP asymmetries of theses decay modes are calculated, some decay modes are first predicted, and some useful relations based on $SU(3)$ flavor symmetry are discussed. Comparing the theoretical results with the current available experimental data, it is found that $K_0^*(1430)$ can be described as the lowest-lying p-wave $(s,u/d)$ state rather than the first excited one.

\end{abstract}

\newpage
\section{Introduction}
The inner structure of scalar~(S) meson has been studied for a long time, and now it is still a hot topic in particle physics. There are several proposals for their inner structure, e.g., $\bar{q}q$ state, tetra-quark states,  molecule states, glueball and hybrid states, while  there is still no general agreement on such issue. For the light scalar mesons, in the 2-quark  scenario~(namely S1), it is suggested  that  the scalar mesons with mass below $1\, {\rm GeV}$~(such as
$a_0(980)$, $f_0(980)$, $\kappa/K_0^*(700)$, etc.) are  interpreted as the lowest lying $q\bar{q}$ states  having a unit of orbital angular momentum and form a $SU(3)$ nonet; while, the ones with mass above  $1\, {\rm GeV}$~(such as $a_0(1450)$, $f_0(1370)$, $K_0^*(1430)$, etc.) are treated as the first  excited $q\bar{q}$ states and are classified into another $SU(3)$ nonet. On the contrary,   in the tetra-quark scenario advocated by Jaffe~\cite{Jaffe:1976ig,Jaffe:1976ih}, the former are  predominately the  $q q \bar{q}\bar{q}$ states without introducing a unit of orbital angular momentum; accordingly, the  later are treated as the lowest lying $q\bar{q}$ p-wave states~(namely S2). This scenario is favored by some lattice calculations~\cite{Mathur:2006bs,Prelovsek:2010kg} and mesonic spectroscopy data~\cite{Close:2002zu}; it is also much more acceptable because the $0^+$ meson has  a unit of orbital angular momentum   and hence should have a higher mass above  $1\, {\rm GeV}$  in the $q\bar{q}$  model. Besides of studies of mass spectra and decays of scalar mesons,  the nonleptonic two-body $B$ meson decay involving a scalar final state, $B\to S M$, provides another efficient way to investigate the features  and  the possible inner structures of scalar mesons.

In recent years, although the identification of scalar mesons is difficult experimentally, some experimental efforts have been devoted to measuring  $B_{u,d}\to S M$ decay modes. For instance, the light scalar $f_0(980)$ was first observed in the $B\to f_0(980) K$ decay by the Belle~\cite{Abe:2002av} and BABAR~\cite{Aubert:2003mi} collaborations in 2002 and 2004, respectively. Since then, more and more $B\to S M$ decay modes have been observed by Belle~\cite{Abe:2006,Belle:2004,Belle:2013bdphi,Belle:2013bdkk}, BABAR~\cite{Lees:2012l,Lees:2011l,BABAR:2007buKK,BABAR:2008,BABAR:2008bdphi,BABAR:2008buphi,BABAR:2009,
BABAR:2011ae,BABAR:2012bdrho} and LHCb~\cite{Aaij:2019loz,Aaij:2019nmr} collaborations. Motivated by the rapid development of experiment, some theoretical studies on these decays are made within some QCD inspired approaches, such as the generalized factorization approach~\cite{Cheng:2020BKKpithr}, QCD factorization (QCDF)~\cite{Cheng:2016Bthr,Cheng:2014Bsthr,Cheng:2013Bthr,Cheng:2007Bthr,Kaminski:2009,Li:2015zra,
Cheng:2005nb,Cheng:2007st,Cheng:2010sn,Cheng:2013fba,Qi:2019qwg}, perturbative QCD approach (PQCD)~\cite{Wang:2019Bsphif0,Zou:2020Bsthr1,Zou:2020Bsthr2,Zou:2017BSD,Li:2019Bkorf,Wang:2014BjpeiKs,Rui:2018,
Rui:2019,Liu:2010Bketa,Liu:2013lka,Liu:2013lxz,Li:2009llw,Li:2019jlp,Zou:2016yhb,Zou:2017zll,Zhang:2013zsl,
Zhang:2012z,Zhang:2010zbud,Zhang:2010zbs,Liu:2021gnp} and other methods~\cite{Cheng:2004cch,Kang:2018a,Aidos:2015a,Cheng:2020a,Li:2013aca,Li:2015rea,Ghahramany:2009gk,
Aliev:2007aas,Yang:2006y,Han:2013hwf,Wang:2015w,Kaminski:2006}. Most of these previous works mainly focus on the $B_{u,d}\to S M$ decay modes, but the $B_s\to S M$ decays have not been fully studied theoretically because most of the $B_s\to S M$ decay modes have not been observed.

In 2019, the $B_s  \to K_0^{*}(1430)^+ K^-+c.c.$ and $B_s  \to \bar K_0^{*}(1430)^0 K^0+c.c.$ decays are observed  for the first time by LHCb collaboration, each with significance over $10$ standard deviations. The measured branching fractions are~\cite{Aaij:2019nmr}
\begin{align}
&{\cal B}(B_s^0 \to K_0^{*}(1430)^{+}K^{-}+c.c.)=(31.3\pm2.3\pm0.7\pm25.1\pm3.3) \times 10^{-6}, \nonumber \\
&{\cal B}(B_s^0 \to \bar K_0^{*}(1430)^0 K^0+c.c.)=(33.0\pm2.5 \pm0.9\pm9.1\pm3.5) \times 10^{-6},
\label{Bskk}
\end{align}
Unfortunately, there is no relevant theoretical prediction for these decays before. It is expected that more $B_s\to S M$ decays will be observed by LHCb and Belle-II collaborations in the near future. Therefore, we would like to make a detailed study of $B_s\to K_0^{*}(1430) P$ and $K_0^{*}(1430) V$~($P$ denotes pseudoscalar meson, and   $V$ denotes vector meson) decay modes within the framework of QCDF in this paper. Besides, the $B_{u,d}\to K_0^{*}(1430)  P$ and $ K_0^{*}(1430) V$  decays will also be investigated~\footnote{The $B_{u,d}\to SP$ and $ SV$ decay modes have been investigated in detail in Ref.~\cite{Cheng:2005nb} and Ref.~\cite{Cheng:2007st}, respectively. Some problems in these works are corrected and  the predictions are updated by the same authors in Ref.~\cite{Cheng:2013fba}. Besides the corrections made in Ref.~\cite{Cheng:2013fba},  some essential  improvements will be made in this work.}.  In order to test whether $K_0^{*}(1430)$ is a lowest lying $q\bar{q}$ state or a first excited state, our calculation and analysis will be made within the just mentioned two scenarios~(S1 and S2).

This paper is organized as follows. In section 2, we present the theoretical framework and calculations for $B_{u,d,s}  \to K_0^*(1430)P,\, K_0^*(1430)V$ decays. In section 3, the values of nonperturbative input parameters are calculated; after that, the numerical results and discussions are presented. Finally, we give our summary in section 4. The definitions of decay constant, form factor and distribution amplitude are given in appendix A, and the amplitudes of  $B_{u,d,s}  \to K_0^*(1430)P,\, K_0^*(1430)V$ decays are summarized in appendix B.

\section{Theoretical framework}
In the Standard Model~(SM), the effective weak Hamiltonian responsible for $B_{u,d,s}\to M_1M_2 $ decay induced by $b\to p$ transition is given as~\cite{Buchalla:1996vs}
\begin{eqnarray}\label{eq:eff}
 {\cal H}_{\rm eff} &=& \frac{G_F}{\sqrt{2}} \biggl[V_{ub}
 V_{up}^* \left(C_1 O_1^u + C_2 O_2^u \right) + V_{cb} V_{cp}^* \left(C_1
 O_1^c + C_2 O_2^c \right)  \nonumber\\
 &&- V_{tb} V_{tp}^*\, \big(\sum_{i = 3}^{10}
 C_i O_i \big. \biggl.  \biggl. \big. + C_{7\gamma} O_{7\gamma} + C_{8g} O_{8g}\big)\biggl] +
 {\rm h.c.},
\end{eqnarray}
where $V_{qb} V_{qp}^{\ast}$~($q=u, c$ and $t$) are products of the Cabibbo-Kobayashi-Maskawa~(CKM) matrix elements, $C_{i}$ the Wilson coefficients, and $O_i$ the relevant effective four-quark operators. To obtain the decay amplitude, the main work is to calculate the hadronic matrix element, $\langle M_1 M_2|O_i| B  \rangle $. In the QCDF,  the hadronic matrix element of each operator can be written as the convolution integrals of the scattering kernel with the distribution amplitudes~(DAs) of the participating mesons~\cite{Beneke:1999br,Beneke:2001mg,Beneke:2003zv},
\begin{align}\label{eq:fe}
  \langle M_1M_2|Q_i|{B}\rangle =& \sum_j F_j^{B\to M_1} f_{M_2}\int \,d y\,{\cal T}_{ij}^{I}(y)\, \varphi_{M_2}(y) + [M_1\leftrightarrow M_2]\,\nonumber\\
&  +f_{B}f_{M_{1}}f_{M_{2}}\int d x \,d y \,d z\,{\cal T}_{i}^{II}(x,y,z)\,\varphi_{M_1}(x)\,\varphi_{M_2}(y)\,\varphi_{B}(z)\,,
\end{align}
where $x\,,y\,,z$  are the momentum fractions; $F_j^{B\to M_1}$ is an appropriate form factor of $B\to M_1$ transition; $f_{B}$ and $f_{M}$ are decay constants of  $B$ and light mesons,  respectively. The definitions of decay constant, form factor and DAs are given in the appendix A. The kernels ${\cal T}_{ij}^{I}(y)$ and ${\cal T}_{i}^{II}(x,y,z)$ in Eq.~\eqref{eq:fe}  are hard-scattering functions and are calculable  in perturbation theory. The former starts at tree level and contains the vertex, penguin corrections  at  next-to-leading order in $\alpha_s$; the later contains the order $\alpha_s$ contributions caused by  hard spectator-scattering  and annihilation topologies.

Applying the factorization formula, one can obtain the amplitudes of $ B_{u,d,s}  \to  K_0^*(1430) P $ and $K_0^*(1430) V$  decay modes, which are collected in the appendix B. In each amplitude, the  quantity $A_{M1M2}$ is the factorized matrix element and can be written as
\begin{align}
A_{M1M2}&=\frac{G_F}{\sqrt 2}\begin{cases}-(m_B^2-m_{M_1}^2)U_0^{B\to M_1}(m_{M_2}^2)f_{M_2} & \text{if} ~M_1M_2=SP\,,\\
(m_B^2-m_{M_1}^2)F_0^{B\to M_1}(m_{M_2}^2)f_{M_2} &\text{if}~M_1M_2=PS\,,\\
2m_{M_2}\epsilon^*\cdot p_B U_1^{B\to M_1}(m_{M_2}^2)f_{M_2} &\text{if}~M_1M_2=SV\,,\\
-2m_{M_1}\epsilon^*\cdot p_B A_0^{B\to M_1}(m_{M_2}^2)f_{M_2}  & \text{if}~M_1M_2=VS\,.
\end{cases}
\end{align}
Here, it has been assumed that the final-state meson $M_1$ carries away  the spectator quark from $B$ meson, and the other one is $M_2$. The  coefficients of flavor operators $\alpha_i^p$ appeared in the amplitudes are expressed in terms of the effective coefficients $a_i^p$ as
\begin{align}\label{eq:atree}
\alpha_1(M_1M_2)&=a_1(M_1M_2),\qquad
\alpha_2(M_1M_2)=a_2(M_1M_2),\\
\alpha_3^p(M_1M_2)&=     \begin{cases} a_3^p(M_1M_2)+a_5^p(M_1M_2) & \text{if}~M_1M_2=PS,SV,VS,\\
a_3^p(M_1M_2)-a_5^p(M_1M_2) & \text{if}~M_1M_2=SP,  \end{cases} \\
\alpha_4^p(M_1M_2)&=\begin{cases} a_4^p(M_1M_2)-r_\chi^{M_2} a_6^p(M_1M_2) & \text{if}~M_1M_2=PS,SP,SV,\\
a_4^p(M_1M_2)+r_\chi^{M_2} a_6^p(M_1M_2) & \text{if}~M_1M_2=VS,\end{cases} \\
\alpha_{3,EW}^p(M_1M_2)&=\begin{cases} a_9^p(M_1M_2)+a_7^p(M_1M_2) & \text{if}~M_1M_2=PS,SV,VS,\\
a_9^p(M_1M_2)-a_7^p(M_1M_2) & \text{if}~M_1M_2=SP,   \end{cases} \\
\label{eq:a4EW}
\alpha_{4,EW}^p(M_1M_2)&=\begin{cases} a_{10}^p(M_1M_2)-r_\chi^{M_2} a_8^p(M_1M_2) &\text{if}~ M_1M_2=PS,SP,SV,\\
a_{10}^p(M_1M_2)+r_\chi^{M_2} a_8^p(M_1M_2) &\text{if}~M_1M_2=VS\,,\end{cases}
\end{align}
where the ratio $r_\chi$ is defined as
\begin{align}
r_{\chi}^P(\mu)=&\frac{2m_P^2}{m_b(\mu)(m_1(\mu)+m_2(\mu))},\qquad
r_\chi^V(\mu)=\frac{2m_V}{m_b(\mu)}\frac{f_V^\perp(\mu)}{f_V},\\
r_{\chi}^{S}(\mu)=&\frac{2m_{S}}{m_b(\mu)}\frac{\bar f_S(\mu)}{f_S}=\frac{2m_{S}^2}{m_b(\mu)(m_1(\mu)-m_2(\mu))}\,.
\end{align}

In Eqs.~(\ref{eq:atree}-\ref{eq:a4EW}), the general form of  effective coefficient $a_i^p$ at  next-to-leading order in $\alpha_s$ is
\begin{align}
a_i^p(M_1M_2)=\left (C_i+\frac{C_{i\pm1}}{N_c}\right)N_i(M_2)+\frac{C_{i\pm1}}{N_c}\frac{C_F \alpha_s}{4\pi}\left[V_i(M_2)+\frac{4\pi^2}{N_c}H_i(M_1M_2)\right]+P_i^p(M_2),
\end{align}
where $i=1,...,10$, the upper~(lower) sign applies when $i$ is odd~(even), $C_F=(N_c^2-1)/(2N_c)$ with $N_c=3$, and $N_i(M_2)=1$ except for $N_{6,8}(V)=0$.  The terms proportional to $N_i(M_2)$ are the leading order contributions, and are the same as the results obtained by naive factorization approach;  the $\alpha_s$ corrections are encoded in the quantities $V_i(M_2)$, $H_i(M_1M_2)$ and $P_i^p(M_2)$, which are obtained by calculating vertex, hard-spectator and penguin diagrams, and can be written as the convolution integrals of the hard-scattering kernels with meson  light-cone DAs. The convolution integrals for these quantities can be evaluated by using expansions of the DAs in terms of Gegenbauer polynomials. In this work, we include the first four terms in the Gegenbauer expansion for the twist-2 DAs due to the nontrivial contribution related to the odd  Gegenbauer moments of scalar meson.

 The integral forms of  $ V_i (M_2)$  and $P_i^p(M_2)$  for $B\to SP$ and $SV$ modes are the same as the ones for  $B\to PP$ and $PV$ modes, which have been obtained in Ref.~\cite{Beneke:2003zv}. After integrating out the momentum fraction, we obtain
\begin{align}
V_i(M_2)=( 12\ln\frac{m_b}{\mu}-\frac{37}{2}- 3i\pi) a_0^{M_2}  +(\frac{11}{2}-3i\pi) a_1^{M_2}-\frac{21}{20} a_2^{M_2}+(\frac{79}{36}-\frac{2i\pi}{3}) a_3^{M_2}
\end{align}
for $i=1-4,9,10$, and
\begin{align}\label{eq:v57}
V_i({M_2})=&(-12\ln\frac{m_b}{\mu}+\frac{13}{2}+3i\pi)a_0^{M_2}    +(\frac{11}{2}-3i\pi)a_1^{M_2}
+\frac{21}{20}a_2^{M_2}+(\frac{79}{36}-\frac{2i\pi}{3})a_3^{M_2}
\end{align}
for $i=5,7$ when $M_2=P\,, S\,, V$; while, for $i=6$ and $8$, we have
\begin{align}
V_i(M)=\begin{cases} -6  &\text{if}~M_2=P~\text{and}~S\,,\\
9-6\pi i  &\text{if}~M_2=V\,. \end{cases}
\end{align}
It is noted that our result given by Eq.~\eqref{eq:v57} is different from the one given in Ref.~\cite{Cheng:2005nb} but is consistent with the one given by the same authors in Ref.~\cite{Cheng:2007st}. For the penguin functions $G_{M_2}(s)$  appeared in $P_i^p(M_2)$~(one may refer to  Ref.~\cite{Beneke:2003zv} for detail), we have the following  analytic results,
\begin{small}
\begin{align}
G_{M_2}(0)=&\left(\frac53 + \frac{2i\pi}{3}\right) a_0^{ {M_2}} + \frac{a_1^{ {M_2}}}{2}
   + \frac{a_2^{ {M_2}}}{5} +\frac{a_3^{ {M_2}}}{9},\\
   G_{M_2}(s_c) =& (\frac53 - \frac23\ln s_c)  a_0^{ {M_2}} + \frac{a_1^{ {M_2}}}{2}
    + \frac{a_2^{ {M_2}}}{5} + \frac{a_3^{ {M_2}}}{9} + \frac43 \left( 8 a_0^{ {M_2}} + 9 a_1^{ {M_2}}
    + 9 a_2^{ {M_2}}+ 9 a_3^{ {M_2}} \right) s_c \nonumber\\
  & + 2(8 a_0^{ {M_2}} + 63 a_1^{ {M_2}} + 214 a_2^{ {M_2}}+ \frac{1505 a_3^{ {M_2}}}{3}) s_c^2
    - 24 (9 a_1^{ {M_2}} + 80 a_2^{ {M_2}}+ \frac{9490 a_3^{ {M_2}}}{27}) s_c^3
    \nonumber\\
     & + (2880 a_2^{ {M_2}} + \frac{91000 a_3^{ {M_2}}}{3}) s_c^4- 39200 a_3^{ {M_2}} s_c^5
  - \frac23\sqrt{1-4s_c}\,\bigg[ (1 + 2 s_c)  a_0^{ {M_2}}
     \nonumber\\
   &+ 6 (4  a_0^{ {M_2}} + 27 a_1^{ {M_2}} + 78 a_2^{ {M_2}}+160  a_3^{ {M_2}}) s_c^2- 36 (9 a_1^{ {M_2}} + 70 a_2^{ {M_2}}+ \frac{2440  a_3^{ {M_2}}}{9}) s_c^3
    \nonumber\\
   &  + (4320 a_2^{ {M_2}} +40600  a_3^{ {M_2}})s_c^4 -58800  a_3^{ {M_2}} s_c^5  \bigg] \left(
    2 \,\mbox{arctanh}\sqrt{1-4 s_c} - i\pi \right)+ 12 s_c^2\,\bigg[ a_0^{ {M_2}}   \nonumber\\
   &+ 3 a_1^{ {M_2}}+ 6 a_2^{ {M_2}}+ 10 a_3^{ {M_2}}
    - \frac43 \left( a_0^{ {M_2}} + 9 a_1^{ {M_2}} + 36 a_2^{ {M_2}}+ 100 a_3^{ {M_2}} \right) s_c+ 18 (a_1^{ {M_2}} + 10 a_2^{ {M_2}}+ 50 a_3^{ {M_2}}) s_c^2
    \nonumber\\
   &
    - (240 a_2^{ {M_2}}+2800 a_3^{ {M_2}}) s_c^3+\frac{9800  a_3^{ {M_2}}}{3}s_c^4 \bigg] \left(
    2 \,\mbox{arctanh}\sqrt{1-4 s_c} - i\pi \right)^2  \,,
\\
   G_{M_2}(1) =& \left(\frac{85}{3} - 6\sqrt3\,\pi + \frac{4\pi^2}{9}\right)  a_0^{ {M_2}}
    - \left( \frac{155}{2} - 36\sqrt3\,\pi+12\pi^2 \right)
    a_1^{ {M_2}} \nonumber\\
   &+ \left( \frac{7001}{5} - 504\sqrt3\,\pi
    + 136\pi^2 \right) a_2^{ {M_2}}-\left( \frac{146581}{9} -6000\sqrt3\,\pi
    + \frac{14920}{9}\pi^2 \right) a_3^{ {M_2}}  \,.
\end{align}
\end{small}
where, $s_c=(m_c/m_b)^2$; while, the functions $\hat{G}_{M_2}(s)$ read
\begin{small}
\begin{align}
\hat G_{M_2}(s_c)=&\frac{16}{9}(1-3s_c)-\frac23 \text{ln} s_c -\frac23(1-4 s_c)^{3/2} (2 \text{arctanh} \sqrt{1-4 s_c} -i \pi) \,, \nonumber \\
\hat G_{M_2}(0)=& \frac{16}{9}+\frac{2\pi}{3}i \,,\quad  \hat G_{M_2}(1)= \frac{2\pi}{\sqrt 3}-\frac{32}{9} \,,
\end{align}
\end{small}
when $M_2=P\,, S$, and
\begin{small}
\begin{align}
\hat G_{M_2}(s_c)=&1-36s_c+12s_c \sqrt{1-4 s_c} (2 \text{arctanh} \sqrt{1-4 s_c} -i \pi)-12s_c^2(2 \text{arctanh} \sqrt{1-4 s_c} -i \pi)^2 \,, \nonumber \\
\hat G_{M_2}(0)=& 1 \,,\quad  \hat G_{M_2}(1)= -35+4\sqrt 3 \pi +\frac{4\pi^2}{3} \, ,
\end{align}
\end{small}
when $M_2=V$.


The  hard-spectator corrections can be written as
\begin{align}
H_i(M_1M_2)=\frac{B_{M_1M_2}}{A_{M_1M_2}}\int_0^1\frac{d\xi}{\xi}\Phi_B(\xi)\int_0^1dx\int_0^1dy
\left[\frac{\Phi_{M_2}(x)\Phi_{M_1}(y)}{\bar x \bar y}\pm\gamma_\chi^{M_1}\frac{\Phi_{M_2}(x)\phi_{M_1}(y)}{ x \bar y} \right]
\end{align}
for $i=1-4,9,10$,
\begin{align}
H_i(M_1M_2)=&-\frac{B_{M_1M_2}}{A_{M_1M_2}}\int_0^1\frac{d\xi}{\xi}\Phi_B(\xi)\int_0^1dx\int_0^1dy
\left[\frac{\Phi_{M_2}(x)\Phi_{M_1}(y)}{ x \bar y}\pm\gamma_\chi^{M_1}\frac{\Phi_{M_2}(x)\phi_{
M_1}(y)}{\bar x \bar y} \right]
\end{align}
for $i=5,7$, and $H_i(M_1M_2)=0$ for $ i=6,8$, where $\bar x=1-x$, $\bar y=1-y$.
The upper~(lower) sign should be applied when $M_1=V\,,P$~($M_1=S$). The quantity $B_{M_1M_2}$ is defined as
\begin{align}
B_{M1M2}&=\frac{G_F}{\sqrt 2}\begin{cases}f_{B_q}f_{M_1}f_{M_2} &\text{if}~M_1M_2=PS,SP,\\
-f_{B_q}f_{M_1}f_{M_2} & \text{if}~M_1M_2=VS,SV.
\end{cases}
\end{align}
Integrating out the momentum fraction, we can finally obtain
\begin{numcases}{H_i=\frac{B_{M_1M_2}}{A_{M_1M_2}}\frac{m_B}{\lambda_B}  \cdot}
 \Big[3 (a_0^{M_1}+a_1^{M_1}+a_2^{M_1}+a_3^{M_1}) \cdot 3 (a_0^{M_2}+a_1^{M_2}+a_2^{M_2}+a_3^{M_2}) \nonumber\\
 ~~\pm  3 \gamma_{\chi}^{M_1}(a_0^{M_2}-a_1^{M_2}+a_2^{M_2}-a_3^{M_2}) X_H  \Big],& $M_1=P~(S)$, \label{eq:110}\\
 \Big[ 3 (a_0^{M_1}+a_1^{M_1}+a_2^{M_1}+a_3^{M_1}) \cdot 3 (a_0^{M_2}+a_1^{M_2}+a_2^{M_2}+a_3^{M_2}) \nonumber\\
 ~~+ 3 \gamma_{\chi}^{M_1}(a_0^{M_2}-a_1^{M_2}+a_2^{M_2}-a_3^{M_2}) (3X_H-6)  \Big],& $M_1=V$,
  \end{numcases}
  for $i=1-4,9,10$, and
\begin{numcases}{H_i=-\frac{B_{M_1M_2}}{A_{M_1M_2}}\frac{m_B}{\lambda_B}  \cdot}
\Big [ 3 (a_0^{M_1}+a_1^{M_1}+a_2^{M_1}+a_3^{M_1}) \cdot 3 (a_0^{M_2}-a_1^{M_2}+a_2^{M_2}-a_3^{M_2}) \nonumber\\
 ~~\pm  3 \gamma_{\chi}^{M_1}(a_0^{M_2}+a_1^{M_2}+a_2^{M_2}+a_3^{M_2}) X_H \Big ], &$M_1=P~(S)$,\label{eq:57}\\
\Big[ 3 (a_0^{M_1}+a_1^{M_1}+a_2^{M_1}+a_3^{M_1})  \cdot 3 (a_0^{M_2}-a_1^{M_2}+a_2^{M_2}-a_3^{M_2}) \nonumber\\
 ~~+ 3 \gamma_{\chi}^{M_1}(a_0^{M_2}+a_1^{M_2}+a_2^{M_2}+a_3^{M_2}) (3X_H-6)  \Big],& $M_1=V$,
 \end{numcases}
 for $i=5$ and $7$,  where, the upper~(lower) sign in Eqs.~(\ref{eq:110}) and ~(\ref{eq:57}) is applied when $M_1=P$~($S$). As has been discussed in many previous works~\cite{Beneke:2003zv,Cheng:2013fba,Cheng:2004cch,Cheng:2009c1,Cheng:2009c2,chang:2014s,chang:2015hu,Sun:2015c,chang:2015s}, hard-spectator corrections suffer from the end-point divergence, which is usually parameterized by the end-point parameter
 \begin{align}\label{eq:XH}
X_H=\ln\left(\frac{m_B}{\Lambda_h}\right)(1+\rho_H e^{i \phi_H})\,.
\end{align}
 The parameters $\rho_H$ and $\phi_H$ reflect the strength and possible strong phase of the end-point contributions, respectively.

The amplitudes of $ B_{u,d,s}  \to  K_0^*(1430) P $ and $K_0^*(1430) V$ decays  collected in the appendix B also receives the contributions of  weak annihilation, which are involved in the  effective coefficients $\beta_i^p$ defined as
\begin{align}
\beta_i^p(M_1M_2)\equiv\frac{B_{M_1M_2}}{A_{M_1M_2}}b_i^p\,,
\end{align}
where,\footnote{ In Refs.~\cite{Cheng:2005nb,Cheng:2007st}, the superscript of the last terms in $b_4^p$ and $b_{3,EW}^p$~(Eq.~(3.12)) should be corrected, {\it i.e.}, $A_2^f \to A_2^i$ and $A_3^i \to A_3^f$.}
\begin{align}
b_1&=\frac{C_F}{N_c^2}C_1A_1^i, &b_2&=\frac{C_F}{N_c^2}C_2A_1^i, \nonumber\\
b_3^p&=\frac{C_F}{N_c^2}[C_3A_1^i+C_5(A_3^i+A_3^f)+N_c C_6 A_3^f],  &b_4^p&=\frac{C_F}{N_c^2}[C_4A_1^i+ C_6 A_2^i],\nonumber \\
b_{3,EW}^p&=\frac{C_F}{N_c^2}[C_9A_1^i+C_7(A_3^i+A_3^f)+N_c C_8 A_3^f], &b_{4,EW}^p&=\frac{C_F}{N_c^2}[C_{10}A_1^i+ C_8 A_2^i].
\end{align}
The subscripts $n=1\,,2\,,3$ of $ A_n^{i,f}$ correspond to the possible Dirac structures $(V-A)(V-A)$, $(V-A)(V+A)$ and $(S-P)(S+P)$, respectively; and the supscripts $i$ and $f$ refer to gluon emission from the initial- and final-state quarks, respectively. The explicit  expressions of $A_n^{i,f}$ for $B\to PS$ and $SP$ decays can be written as
\begin{align}
A_1^i&=\pi\alpha_s\int_0^1dxdy  \left\{ \Phi_{M_2}(x)\Phi_{M_1}(y)\left[\frac{1}{y(1-x\bar y)}+\frac{1}{\bar x^2 y}\right] \mp \gamma_\chi^{M_1}\gamma_\chi^{M_2}\phi_{M_2}(x)\phi_{M_1}(y)\frac{2}{\bar x y}\right\}\,, \\
 A_1^f&=0\,;   \\
 A_2^i&=\pi\alpha_s\int_0^1dxdy \left\{   - \Phi_{M_2}(x)\Phi_{M_1}(y)\left[\frac{1}{\bar x(1-x\bar y)}+\frac{1}{\bar x y^2}\right]\pm \gamma_\chi^{M_1}\gamma_\chi^{M_2}\phi_{M_2}(x)\phi_{M_1}(y)\frac{2}{\bar x y}\right\}\,,  \\
 A_2^f&=0,    \\
 A_3^i&=\pi\alpha_s\int_0^1dxdy \left\{  \pm \gamma_\chi^{M_1}\Phi_{M_2}(x)\phi_{M_1}(y) \frac{2\bar y}{\bar x y(1-x\bar y)} + \gamma_\chi^{M_2}\phi_{M_2}(x)\Phi_{M_1}(y)\frac{2x}{\bar x y(1-x\bar y)} \right\} \,, \\
 A_3^f&=\pi\alpha_s\int_0^1dxdy \left\{  \pm  \gamma_\chi^{M_1}\Phi_{M_2}(x)\phi_{M_1}(y) \frac{2(1+\bar x)}{\bar x^2 y} - \gamma_\chi^{M_2}\phi_{M_2}(x)\Phi_{M_1}(y)\frac{2(1+y)}{\bar x y^2} \right\} \,,
 \end{align}
where, the upper and lower signs are applied to $B\to PS$ and $SP$ decays, respectively. The $A_n^{i,f}$ for $B\to VS$ decay can be obtained from the $A_n^{i,f}$ for $B\to PS$ decay by changing the sign of second term in $A_{1,3}^i$ and  $A_3^f$, and the first term in  $A_2^i$; while, the $A_n^{i,f}$ for $B\to SV$ decay is the same as $A_n^{i,f}$ for $B\to SP$ decay except for the overall sign of $A_2^i$.
After integrating out the momentum fractions, we can finally obtain
\begin{align}
A_1^i(PS) &\approx  2\pi \alpha_s \bigg\{9\bigg[ a_0^{ {M_2}}(X_A-4+\frac{\pi^2}{3})+ a_1^{ {M_2}}(3X_A+4-\pi^2)+ a_2^{ {M_2}}(6X_A-\frac{107}{3}+2\pi^2)\nonumber \\
&+a_3^{ {M_2}} (10X_A+\frac{23}{18}-\frac{10}{3}\pi^2)\bigg]- \gamma_\chi^P \gamma_\chi^S X_A^2 \bigg\}, \\
A_2^i(PS) &\approx  2\pi \alpha_s \bigg\{-9\bigg[a_0^{ {M_2}}(X_A-4+\frac{\pi^2}{3}) + a_1^{ {M_2}}(X_A+29-3\pi^2)+a_2^{ {M_2}}(X_A-119+12\pi^2)\nonumber\\
&+a_3^{ {M_2}}(X_A+\frac{2956}{9}-\frac{100}{3}\pi^2 )\bigg]+\gamma_\chi^P \gamma_\chi^S  X_A^2 \bigg\}, \\
A_3^i(PS) &\approx   6\pi \alpha_s \bigg\{\gamma_\chi^P \bigg[a_0^{ {M_2}} (X_A^2-2X_A+\frac{\pi^2}{3})+ 3 a_1^{ {M_2}}(X_A^2-4X_A+4+\frac{\pi^2}{3})\nonumber\\
&+ 6 a_2^{ {M_2}}(X_A^2-\frac{16}{3}X_A+\frac{15}{2}+\frac{\pi^2}{3})
+10 a_3^{ {M_2}}(X_A^2-\frac{19}{3}X_A+\frac{191}{18}+\frac{\pi^2}{3}
 )\bigg] \nonumber\\
& +\gamma_\chi^S ( X_A^2-2X_A+\frac{\pi^2}{3}) \bigg\}, \\
A_3^f(PS) &\approx   6\pi \alpha_s X_A \bigg\{\gamma_\chi^P \bigg[a_0^{ {M_2}}(2X_A-1)+ a_1^{ {M_2}}(6X_A-11)+a_2^{ {M_2}}(12X_A-31)\nonumber\\
& +a_3^{ {M_2}}(20X_A-\frac{187}{3} )\bigg]-\gamma_\chi^S( 2X_A-1) \bigg\}
\end{align}
for $ M_1M_2=PS $, and
\begin{align}
A_1^i(VS) &\approx  6\pi \alpha_s \bigg\{3\bigg[ a_0^{ {M_2}}(X_A-4+\frac{\pi^2}{3})+ a_1^{ {M_2}}(3X_A+4-\pi^2)+a_2^{ {M_2}}(6X_A-\frac{107}{3}+2\pi^2)
\nonumber\\
&+a_3^{ {M_2}} (10X_A+\frac{23}{18}-\frac{10}{3}\pi^2 )\bigg]-\gamma_\chi^V\gamma_\chi^S  X_A(X_A-2) \bigg\}, \\
A_2^i(VS) &\approx  6\pi \alpha_s \bigg\{3\bigg[ a_0^{ {M_2}}(X_A-4+\frac{\pi^2}{3})+ a_1^{ {M_2}}(X_A+29-3\pi^2)+a_2^{ {M_2}}(X_A-119+12\pi^2)\nonumber\\
&+a_3^{ {M_2}} (X_A+\frac{2956}{9}-\frac{100}{3}\pi^2 )\bigg]-\gamma_\chi^V\gamma_\chi^S  X_A(X_A-2) \bigg\}, \\
A_3^i(VS) &\approx  6\pi \alpha_s \bigg\{-\gamma_\chi^V \bigg[3 a_0^{ {M_2}}(X_A^2-2X_A+4-\frac{\pi^2}{3}) +9 a_1^{ {M_2}} (X_A^2-4X_A-4+\pi^2  )\nonumber \\
&+3 a_2^{ {M_2}} (6X_A^2-32X_A+79-2\pi^2)+10 a_3^{ {M_2}} (3X_A^2-19X_A+\frac{61}{6} +3 \pi^2 )\bigg]
 \nonumber\\
& -\gamma_\chi^S ( X_A^2-2X_A+\frac{\pi^2}{3} ) \bigg\}, \\
A_3^f(VS) &\approx  6\pi \alpha_s \bigg\{-3\gamma_\chi^V (X_A-2)\bigg[ a_0^{ {M_2}}(2X_A-1) +a_1^{ {M_2}}(6X_A-11)+a_2^{ {M_2}}(12X_A-31)\nonumber\\
& +a_3^{ {M_2}} (20X_A-\frac{187}{3} )\bigg]+\gamma_\chi^SX_A( 2X_A-1) \bigg\}
\end{align}
for $ M_1M_2=VS $, where, $X_A$ is the endpoint parameter and is defined in the same manner as $X_H$ given by Eq.~\eqref{eq:XH}. The results for the cases of $ M_1M_2=SP $ and $ M_1M_2=SV $  can be obtained via the relation
\begin{align}
A_1^i(SP)&=A_2^i(PS),~ A_2^i(SP)=A_1^i(PS),~A_3^i(SP)=-A_3^i(PS), ~A_3^f(SP)=A_3^f(PS),\\
A_1^i(SV)&=-A_2^i(VS),~A_2^i(SV)=-A_1^i(VS),~A_3^i(SV)=A_3^i(VS), ~A_3^f(SV)=-A_3^f(VS)\,,
\end{align}
but with different sign of $a_0^{ {M_2}}$ and $a_2^{ {M_2}}$. In above evaluation, the asymptotic DAs of $P$ and $V$ mesons are used for simplicity; while, for the DA of $S$ meson, the first four terms in Gegenbauer expansion are included because of the dominance of  $a_1^{S}$.  In the previous works~\cite{Cheng:2005nb,Cheng:2007st,Cheng:2010sn,Cheng:2013fba}, the contributions related to  $a_0^{S}$ and $a_2^{S}$ are neglected, which is reasonable for the cases of scalar $(u,d)$ state and quarkonium  because $a_{0,2}^{S}\simeq 0$. However, for the case of $K^*_0(1430)$, the contributions related to non-zero $a_{0,2}^{S}$ are possibly nontrivial compared with the ones  related to $a_{1,3}^{S}$, and hence are considered in this work.

\section{Numerical Results and Discussions}
Before presenting our numerical results, we would like to clarify the input parameters used in the evaluations.  For the CKM matrix elements, we adopt the Wolfenstein  parameterization and choose  the four parameters as~\cite{Zyla:2020p}
  \begin{align}
  A=0.790^{+0.017}_{-0.012}\,, \quad
  \lambda=0.22650^{+0.00048}_{-0.00048}\,,\quad
 \bar{\rho}=0.141^{+0.016}_{-0.017}, \quad
  \bar{\eta}=0.357^{+0.011}_{-0.011}.
  \label{eq:ckminput}
  \end{align}
As for the quark masses, we take~\cite{Zyla:2020p}
  \begin{align}
 &m_{s}(\mu)/m_{q}(\mu)=27.3^{+0.7}_{-1.3}\,,  \quad
 m_{s}(2{\rm GeV})=93^{+11}_{-5}\, {\rm MeV }\,, \quad
 m_{b}(m_b)=4.18^{+0.03}_{-0.02}\, {\rm GeV}\,, \\
&  m_{c}=1.67\pm 0.07 \,  {\rm GeV}\,, \quad
 m_{b}=4.78\pm 0.06\,  {\rm GeV} \,, \quad
  m_{t}=172.76\pm 0.30\,  {\rm GeV}\,,
  \end{align}
where $m_{q}\equiv(m_u+m_d)/2$. For the well-determined Fermi coupling constant, masses of mesons and lifetimes of $B$ mesons, we take their default values given by PDG~\cite{Zyla:2020p}.

The nonperturbative inputs used in this work include decay constant, Gegenbauer moment and form factor. Unfortunately, for the $ B_{u,d,s}  \to K_0^*(1430) P $ and $K_0^*(1430) V$  decays concerned in this work, some of these nonperturbative inputs are not known.  {The standard light-front~(SLF) approach~\cite{Terentev:1976jk,Berestetsky:1977zk,Jaus:1989au,Jaus:1989av}  provides a conceptually simple and phenomenologically feasible framework for calculating  the non-perturbative quantities of hadrons. However, it is powerless for determining the zero-mode contributions by itself and the  Lorentz covariance is lost.  In order to cover the shortages of SLF approach, a manifestly covariant light-front~(CLF) approach is exploited~\cite{Jaus:1999zv,Jaus:2002sv,Cheng:2004cch} with the help of the manifestly covariant Bethe-Salpeter~(BS) approach, and has been applied to study the   $B_{u,d}\to SP$ and $SV$ decays~\cite{Cheng:2005nb,Cheng:2007st,Cheng:2010sn,Cheng:2013fba}. Unfortunately, this traditional  CLF approach  has some self-consistence problems, and  the covariance in fact can not be strictly guaranteed due to the residual  spurious $\omega$-dependent contribution~\cite{Cheng:2004cch,Choi:2013mda,Chang:2019obq}.  In order to  resolve these problems, a self-consistent scheme is presented in Ref.~\cite{Choi:2013mda} by improving the correspondence between CLF and BS calculation, and has been tested in, for instance, Refs.~\cite{Choi:2017zxn,Choi:2017uos,Choi:2021mni,Chang:2018zjq,Chang:2019mmh,Chang:2019obq,Chang:2020wvs,Choi:2021qza,Choi:2020xsr}. Most of the results based on such improved self-consistent CLF approach for the decay constant and form factors generally agree with the experimental data and the predictions obtained by using Lattice QCD~(LQCD) and light cone sum rules (LCSR) (some examples can be found in Refs.~\cite{Chang:2018zjq,Chang:2019mmh,Choi:2021qza,Choi:2021mni,Choi:2017zxn}) , while the self-consistent CLF results for  some DAs of light $P$-mesons~( for instance, the twist-3 DAs of $\pi$ and $K$ mesons~\cite{Choi:2017uos,Choi:2014ifm}) are different from the QCD sum rule~(QCD SR)  results.   }


In the previous works for $B_{u,d}\to SP$ and $SV$ decays~\cite{Cheng:2005nb,Cheng:2007st,Cheng:2010sn,Cheng:2013fba}, the decay constant and Gegenbauer moments are evaluated by using the QCD SR, while and the form factor is  evaluated by using the traditional CLF approach. In this work, all of the  nonperturbative parameters will be calculated by using the self-consistent CLF approach for consistence. The theoretical framework for the decay constant of $S$, $P$ and $V$  mesons and the form factors of $P\to (S\,,P\,,V)$  transitions within the CLF approach has been given in, for instance, Refs.~\cite{Cheng:2004cch,Chang:2018zjq}, and the self-consistent correspondence relation between the BS and the LF approaches has been discussed in detail in Refs.~\cite{Choi:2013mda,Choi:2017zxn,Choi:2017uos,Choi:2021mni,Chang:2018zjq,Chang:2019mmh,Chang:2019obq,Chang:2020wvs}. Using the Gaussian-type wavefunctions~\footnote{Our sign convention for $\psi_{2p}(x,\kb)$ is different from the one in Ref.~\cite{Hwang:2010iq}, and ensures the positive decay constant of $S(2P)$ and form factor of $P\to S(2P)$ transition. It should be noted that the different conventions do not affect the final results of observables. }
\begin{eqnarray}\label{eq:GW1}
\psi_{1s}(x,\kb) &=&\frac{4\pi^{\frac{3}{4}}}{\beta^{\frac{3}{2}}} \sqrt{ \frac{\partial k_z}{\partial x}}\exp\left[ -\frac{\overrightarrow{k}^2}{2\beta^2}\right]\,,\\
\psi_{1p}(x,\kb) &=&\frac{\sqrt{2}}{\beta}\psi_{1s}(x,\kb) \,,\\
\psi_{2p}(x,\kb) &=&\sqrt{\frac{5}{2}}\left( \frac{2\overrightarrow{k}^2}{5\b^2}-1   \right)\psi_{1p}(x,\kb) \,,
\label{eq:GW3}
\end{eqnarray}
and the values of input parameters collected in Ref.~\cite{Chang:2019obq}, we obtain~(in units of MeV)
\begin{align}
&f_{K_0^*(1430)}=18\pm5    \qquad   \text{S1},  \qquad \qquad \,\,\, f_{K_0^*(1430)}=43\pm10    \qquad   \text{S2},\\
&f_{B_{u,d}}=186\pm7,\qquad\quad f_{B_s}=224\pm9,\qquad\quad f_{\pi}=131\pm7,\qquad\quad \,f_{K}=156\pm5, \\
&f_{K^*}=205\pm8,\qquad\quad \, \,f_{\rho}=210\pm4,\qquad\quad \,\,~f_{\omega}=210\pm4,\qquad\quad \,f_{\phi}=228\pm5, \\
&f_{K^*}^{\perp}=173\pm6,\qquad\quad \, f_{\rho}^{\perp}=167\pm4,\qquad\quad \,~f_{\omega}^{\perp}=167\pm4,\qquad\quad f_{\phi}^{\perp}=198\pm4,
\end{align}
where,  as has been mentioned in the introduction, S1 and S2 are the two assumptions for $K_0^*(1430)$ meson that
\begin{itemize}
\item S1: $K_0^*(1430)$ is the first excited p-wave two-quark state;
\item S2: $K_0^*(1430)$ is the lowest-lying p-wave two-quark state.
\end{itemize}
 It is found that our result for $f_{K_0^*(1430)}$ in S2 is in agreement with the results $f_{K_0^*}=34\pm7\,\rm MeV$~\cite{Du:2005dly} {and $42\pm8\,\rm MeV$~\cite{Yang:2005bu} obtained within QCD SR} and $f_{K_0^*}=42\pm2\,\rm MeV$~\cite{Maltman:1991k} obtained within the finite-energy sum rules.  {The heavy lepton $\tau$ decay process $\tau\to K^*_0(1430)\nu_\tau$ is very suitable for testing the value of $f_{K_0^*}$ because there is only one hadron evolved in this decay, the  interaction in the leptonic vertex can be calculated with high precision, and its branching fraction is dependent on $f_{K_0^*}$ directly. Using our prediction $f_{K_0^*}=43 \,{\rm MeV}$ and the formulae given in Ref.~\cite{Yang:2005bu}, we obtain ${\cal B}(\tau\to K^*_0(1430)\nu) \simeq 0.81\times 10^{-4}$~(S2), which is allowed by current data ${\cal B}(\tau\to K^*_0(1430)\nu)^{\rm exp.}<5\times 10^{-4}$~\cite{Zyla:2020p}.
 }


 \begin{table}[t]
\caption{Form factors of $B\to K_0^*(1430),K, K^*, \pi,  \rho, \omega$ and  $B_s\to K_0^*(1430),K,  K^*, \phi$ transitions  in S1 (lower entry) and S2 (upper entry) by using the self-consistent CLF approach.}
\vspace{-0.1cm}
\begin{center}\setlength{\tabcolsep}{5pt}
\begin{tabular}{llllllrl}
\hline\hline
\quad F&$\quad F(0)$&\,\,\,a&\,\,\,b&               \quad F&$ \,\,\,F(0)$&a&\,\,\,b    \\  \hline
  $U_1^{B\to K_0^*}$&$0.29\pm0.02$&1.27&0.33&
  $U_0^{B\to K_0^*}$&$0.29\pm0.02$&0.16&0.11\\
                      &$0.18\pm0.01$&1.03&0.15& &$0.18\pm0.01$&$-0.23$&0.29\\
 $U_1^{B_s\to K_0^*}$&$0.28\pm0.02$&1.58&0.84&
 $U_0^{B_s\to K_0^*}$&$0.28\pm0.02$&0.55&0.20\\
                      &$0.23\pm0.02$&0.92&0.29& &$0.23\pm0.02$&$-0.23$&0.36\\
$F_0^{B\to\pi}$&$0.27\pm0.03$&0.65&0.03&
$F_0^{B\to K}$&$0.34\pm0.03$&0.66&0.07
\\
 $F_0^{B_s\to K}$&$0.23\pm0.03$&0.95&0.28&
 $A_0^{B\to \omega}$&$0.29\pm0.03$&1.54&0.74
\\
 $A_0^{B\to\rho}$&$0.29\pm0.03$&1.54&0.74&
 $A_0^{B\to K^*}$&$0.34\pm0.04$&1.52&0.65
 \\
 $A_0^{B_s\to K^*}$&$0.21\pm0.04$&1.94&1.62&
 $A_0^{B_s\to \phi}$&$0.28\pm0.04$&1.80&1.29
 \\
  \hline\hline
\end{tabular}
\end{center}
\label{tab:BSPSVFF}
\end{table}

\begin{table}[t]
{\caption{Form factors of $B\to K, K^*, \pi,  \rho, \omega$ and  $B_s\to K,  K^*, \phi$ transitions at $q^2=0\,{\rm GeV}^2$ obtained in this and some previous works .}}
\vspace{-0.1cm}
\begin{center}\setlength{\tabcolsep}{5pt}
\begin{tabular}{llcccccccccc}
\hline\hline
 &This work&LCSR&SCET&LFQM&pQCD&CCQM\\
 & &\cite{Ball:2004ye,Ball:2004rg}&\cite{Lu:2007sg}&\cite{Verma:2011yw}&\cite{Wang:2012ab,Li:2009tx}&\cite{Soni:2020bvu}\\  \hline
 $F_0^{B\to\pi}$  &$0.27\pm0.03$&$0.258\pm0.030$&0.247&$0.25$&$0.26^{+0.05}_{-0.05}$&$0.283 \pm 0.019$\\
 $F_0^{B\to K}$   &$0.34\pm0.03$&$0.331\pm0.040$&0.297&$0.34$&$0.31^{+0.05}_{-0.05}$&---\\
 $F_0^{B_s\to K}$&$0.23\pm0.03$&---&0.290&$0.23$&$0.26^{+0.05}_{-0.05}$& $0.247 \pm 0.015$ \\ \hline \hline

 $A_0^{B\to\rho}$&$0.29\pm0.03$&$0.303\pm0.028$&0.260&$0.32$&$0.25^{+0.07}_{-0.06}$&$0.266\pm0.013$\\
 $A_0^{B\to \omega}$&$0.29\pm0.03$&$0.281\pm0.030$&0.240&$0.28$&$0.23^{+0.06}_{-0.05}$&$0.236\pm0.011$\\
 $A_0^{B\to K^*}$&$0.34\pm0.04$&$0.374\pm0.034$&0.283&$0.38$&$0.31^{+0.09}_{-0.07}$&---\\

 $A_0^{B_s\to K^*}$&$0.21\pm0.04$&$0.363\pm0.034$&0.279&$0.25$&$0.24^{+0.06}_{-0.05}$ &$0.225 \pm 0.090$\\
 $A_0^{B_s\to \phi}$&$0.28\pm0.04$&$0.474\pm0.037$&0.279&$0.31$&$0.31^{+0.09}_{-0.07}$&---\\
  \hline\hline
\end{tabular}
\end{center}
\label{tab:B2PVFF}
\end{table}

 Our results for the form factors of $B\to K_0^*(1430),K, K^*, \pi,  \rho, \omega$ and  $B_s\to K_0^*(1430),K,  K^*, \phi$ transitions are collected in Table~\ref{tab:BSPSVFF}.  {Our results $U_{0,1}^{B\to K_0^*}=$$0.29\pm0.02$(S2) and $0.18\pm0.01$(S1) are a little larger and smaller than the traditional CLF results $U_{0,1}^{B\to K_0^*}=$$0.26$(S2) and $0.21$(S1)~\cite{Cheng:2005nb}. The results for $B_s\to K_0^*$ transition are first obtained in this work. The previous results for the $B\to K, K^*, \pi,  \rho, \omega$ and  $B_s\to K,  K^*, \phi$ transitions at $q^2=0\,{\rm GeV}^2$ obtained via  other approaches are collected in Table~\ref{tab:B2PVFF} for comparison.  It can be found that these results based on different approaches are generally consistent with each other except for the large  $A_0^{B_s\to (K^*,\phi)}=(0.363\pm0.034\,,0.474\pm0.037)$ predicted by LCSR~\cite{Ball:2004ye,Ball:2004rg} (relatively small results $A_0^{B_s\to (K^*,\phi)}=(0.314\pm0.048\,,0.389\pm0.045)$ are obtained in Ref.~\cite{Bharucha:2015bzk}).   From above discussion and Table~\ref{tab:B2PVFF}, we can roughly conclude that the uncertainties of form factors associated with  $B$ and $B_s$ decays caused by different approaches are less than about $20\%$ and $35\%$, respectively. }
\begin{table}[t]
\caption{{Gegenbauer moments of  pseudoscalar and vector mesons at $\mu=1\rm GeV$.}}
\vspace{-0.1cm}
\begin{center}
\let\oldarraystretch=\arraystretch
\renewcommand*{\arraystretch}{0.8}
\setlength{\tabcolsep}{8pt}
\begin{tabular}{clcccccccccccc}
\hline\hline
  &                                & $a_1$  &$a_2$  & $a_3$   \\  \hline
  $\pi$  & this work                            &$0$          &$0.10\pm0.04$  &$0$\\
                        &LQCD\cite{RQCD:2019osh}&$0$       &$0.116\pm0.020$ &$0$\\
                        &QCD SR\cite{Ball:2006wn}&$0$          & $0.25\pm 0.15$ &$0$\\
          &linear(HO)\cite{Choi:2007yu}&$0$        &$0.12(0.05)$&$0$\\

  \hline

$\bar K$ & this work             &$0.09\pm0.01$       &$0.02\pm0.02$   &$0.05\pm0.01$\\
   &LQCD\cite{RQCD:2019osh} &$0.053^{+0.031}_{-0.033}$       &$0.106\pm0.016$       &---\\
   &QCD SR\cite{Ball:2006wn}  &$0.06\pm 0.03$ &$0.25\pm 0.15$   &---\\
  &linear(HO)\cite{Choi:2007yu}  &$0.09(0.13)$    &$0.03(-0.03)$    &$0.06(0.04)$\\\hline

  $\rho$& this work   &$0$ &$-0.01\pm0.02$ &$0$\\
    &QCD SR\cite{Ball:2004rg}      &$0$ &$0.09^{+0.10}_{-0.07}$ &$0$\\
  &QCD SR\cite{Ball:2007rt}&$0$&\,\,\,\,$0.15\pm0.07$&$0$\\
  &linear(HO)\cite{Choi:2007yu}&$0$&\,\,\,\,$0.02(-0.02)$&$0$\\\hline

  $\bar K^*$& this work                  &$0.13\pm0.04$  &$-0.07\pm0.00$  &$0.02\pm0.00$\\
  &QCD SR\cite{Ball:2004rg}      &$0.10\pm 0.07$ &$0.07^{+0.09}_{-0.07}$ &---\\
  &QCD SR\cite{Ball:2007rt}      &$0.03\pm 0.02$ &$0.11\pm 0.09$  &---\\
  &linear(HO)\cite{Choi:2007yu}&$0.11(0.14)$     &$-0.03(-0.07)$    &$0.03(0.02)$\\\hline

  $\phi$&this work&$0$ &$-0.13\pm0.03$ &$0$\\
    &QCD SR\cite{Ball:2004rg}     &$0$ &$0.06^{+0.09}_{-0.07}$ &$0$\\
  &QCD SR\cite{Ball:2007rt}   &$0$ &$0.18\pm 0.08$ &$0$\\
   \hline\hline
\end{tabular}
\end{center}
\label{tab:Gegenbauer1}
\end{table}
\begin{table}[t]
\caption{{Gegenbauer moments of  $ {K}_0^*(1430)$ at $\mu=1\rm GeV$ in S1 (lower entry) and S2 (upper entry)}.}
\vspace{-0.1cm}
\begin{center}
\let\oldarraystretch=\arraystretch
\renewcommand*{\arraystretch}{0.8}
\setlength{\tabcolsep}{8pt}
\begin{tabular}{cccccccccccccccccccc}
\hline\hline
                 & $a_1$  &$a_2$  & $a_3$   \\  \hline
this work &$-1.65\pm0.38$&$-0.38\pm0.02$ &$-0.06\pm0.01$\\
                &$-1.93\pm0.69$&$0.75\pm0.13$ &$-1.13\pm0.72$\\\hline
QCD SR~\cite{Cheng:2005nb} &$-7.21\pm1.64$&--- &$-5.31\pm2.70$\\
       &$7.33\pm0.89$&--- &$-15.17\pm1.01$\\
  \hline\hline
\end{tabular}
\end{center}
\label{tab:Gegenbauer2}
\end{table}

\begin{figure}[t]
\begin{center}
\subfigure[]{\includegraphics[scale=0.5]{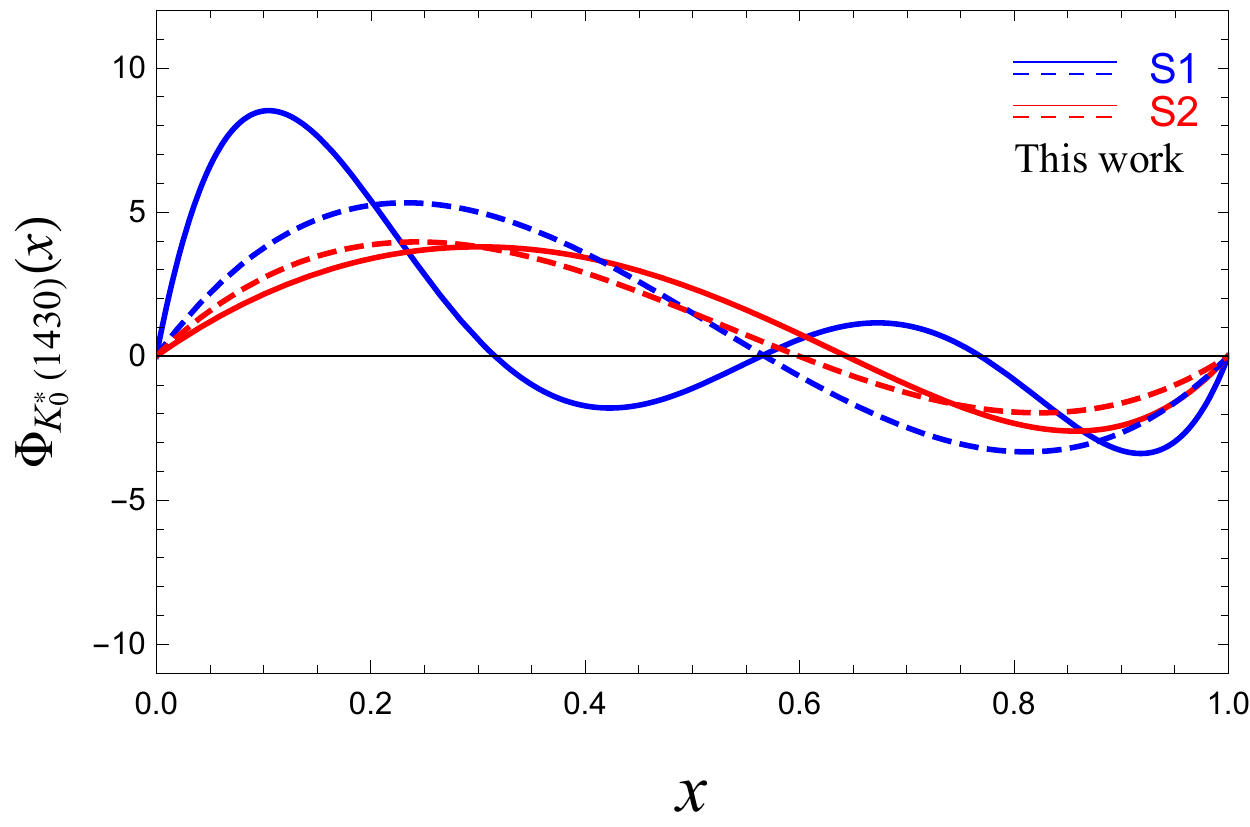}}\qquad
\subfigure[]{\includegraphics[scale=0.5]{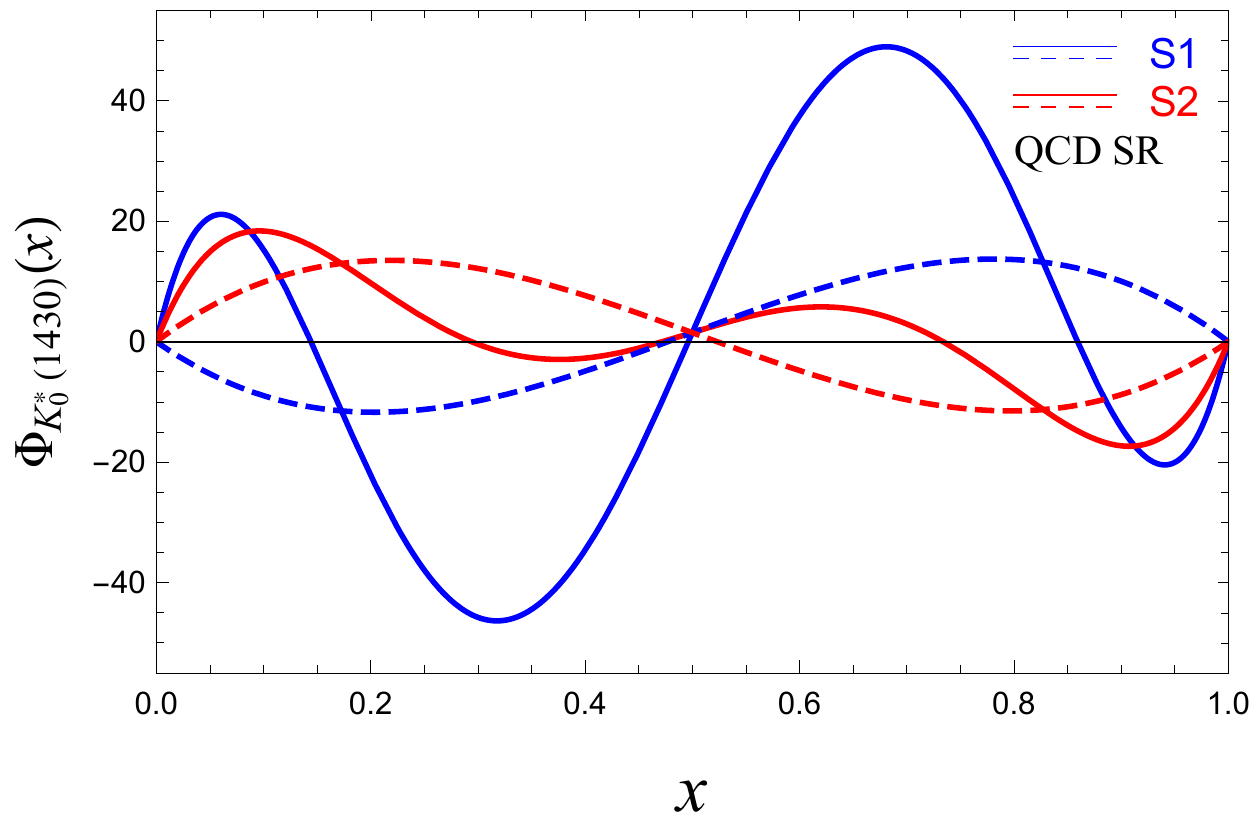}}
\caption{{The DA of $ {K}_0^*(1430)$ predicted in this work and QCD SR~\cite{Cheng:2005nb}. The dashed  and  solid lines correspond to the truncations up to $n=1$ and $n=3$, respectively, according to Eq.~\eqref{eq:tw2Phi}. }   }
 \label{fig:Bphi}
\end{center}
\end{figure}

 The DAs of mesons have been studied within the LF approaches in, for instance, Refs.~\cite{Choi:2013mda,Choi:2014ifm,Choi:2017uos}. The Gegenbauer moments can be extracted from DAs via
\begin{align}
a_n^M=\frac{2(2n+3)}{3(n+1)(n+2)}\int_0^1 {\rm d} x \,C_n^{3/2}(x-\bar{x})\Phi(x)\,,
\end{align}
where, $C_n^{3/2}$ are Gegenbauer polynomials. Our numerical results  for the  Gegenbauer moments at $\mu=1\rm GeV$ are collected {in Tables~\ref{tab:Gegenbauer1} and \ref{tab:Gegenbauer2}. The results obtained by using LQCD~\cite{RQCD:2019osh}, QCD SR~\cite{Ball:2006wn,Ball:2007rt,Ball:2004rg,Cheng:2005nb} and traditional LFQM with  linear confining  and harmonic oscillator (HO) potentials~\cite{Choi:2007yu} are also collected  in these tables for comparison.  For the pseudoscalar and vector mesons, our results are generally agree with the results based on the other approaches within errors except for the signs of $a_2^{\rho,K^*,\phi}$, which is caused by the fact that the DAs predicted by LF approaches are usually different from the QCD SR predictions especially at $x\to 0$ or $1$~(some examples can be found in, for instance, Refs.~\cite{Choi:2017uos,Choi:2007yu,Chang:2016ouf}).  For the  $ {K}_0^*(1430)$ meson,  it is found from Table~\ref{tab:Gegenbauer2} that our results are much smaller than the ones given by QCD SR~\cite{Cheng:2005nb}. In order to clearly show their difference, we plot the DA of $ {K}_0^*(1430)$ in Fig.~\ref{fig:Bphi}. For the case of $n=1$~(dashed lines), our results in S1 and S2 are similar, while the QCD SR  results  in S1 and S2 are totally different with each other due to the different signs of $a_1^{{K}_0^*}$~(QCD SR, $-7.21$ vs. $7.33$). In addition, the $a_3^{{K}_0^*}$ correction to $\Phi_{{K}_0^*}(x)$ in our evaluation is not as significant as the one predicted by QCD SR. More theoretical and experimental efforts are needed for a clear picture of $\Phi_{{K}_0^*}(x)$.~\footnote{{The $\gamma\gamma\to K^{*+}_0(1430) K^{*-}_0(1430)\to 2K2\pi$ process may provide  a way to test the DA of $K^*_0(1430)$, which is similar to the cases of $\gamma\gamma\to \pi^+\pi^-$ and $K^+K^-$ processes for testing the DAs of $\pi$ and $K$ mesons~\cite{Wang:2015mod,Wang:2019trq}. Such process is in the scope of Belle(-II)  in principle~\cite{Belle:2007qae}, but the specific measurement has not been made. } }
 }


\begin{figure}[t]
\begin{center}
\subfigure[]{\includegraphics[scale=0.4]{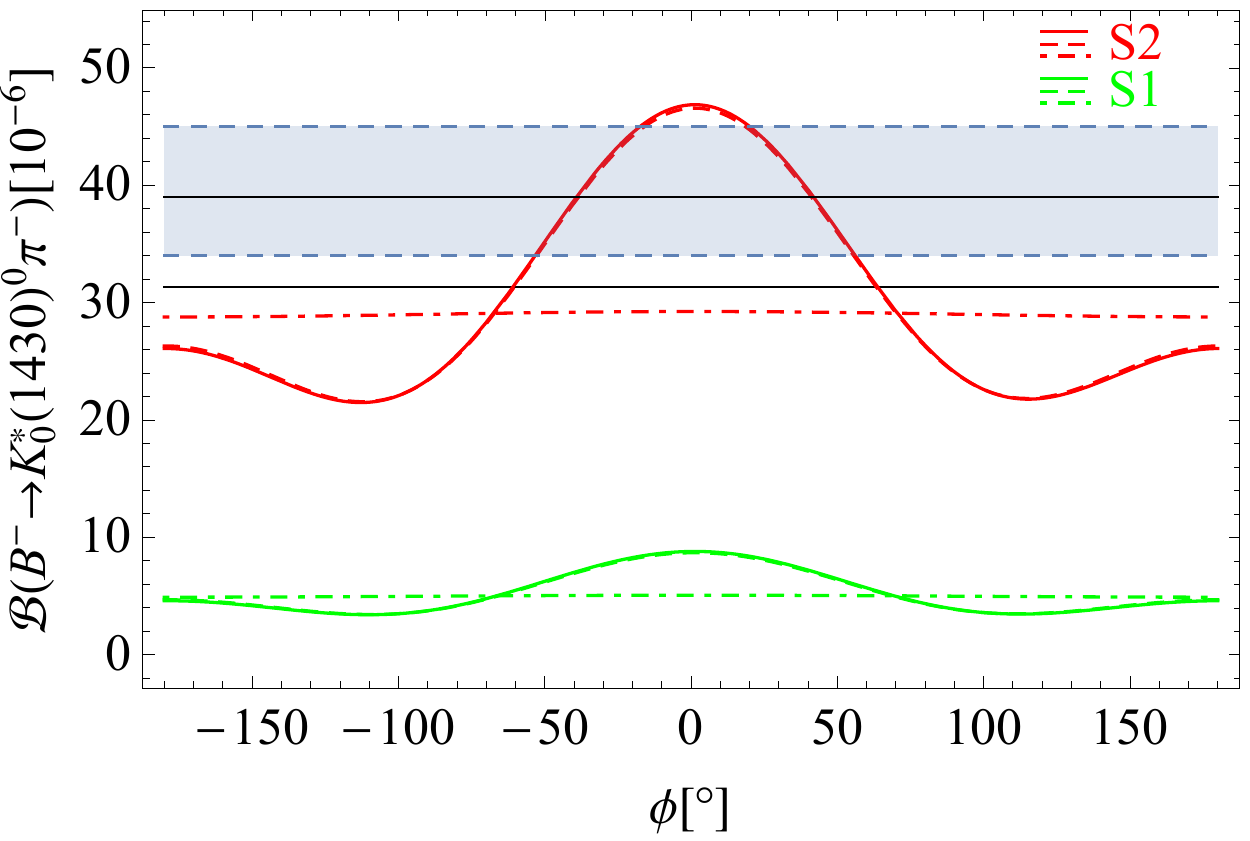}}\quad
\subfigure[]{\includegraphics[scale=0.4]{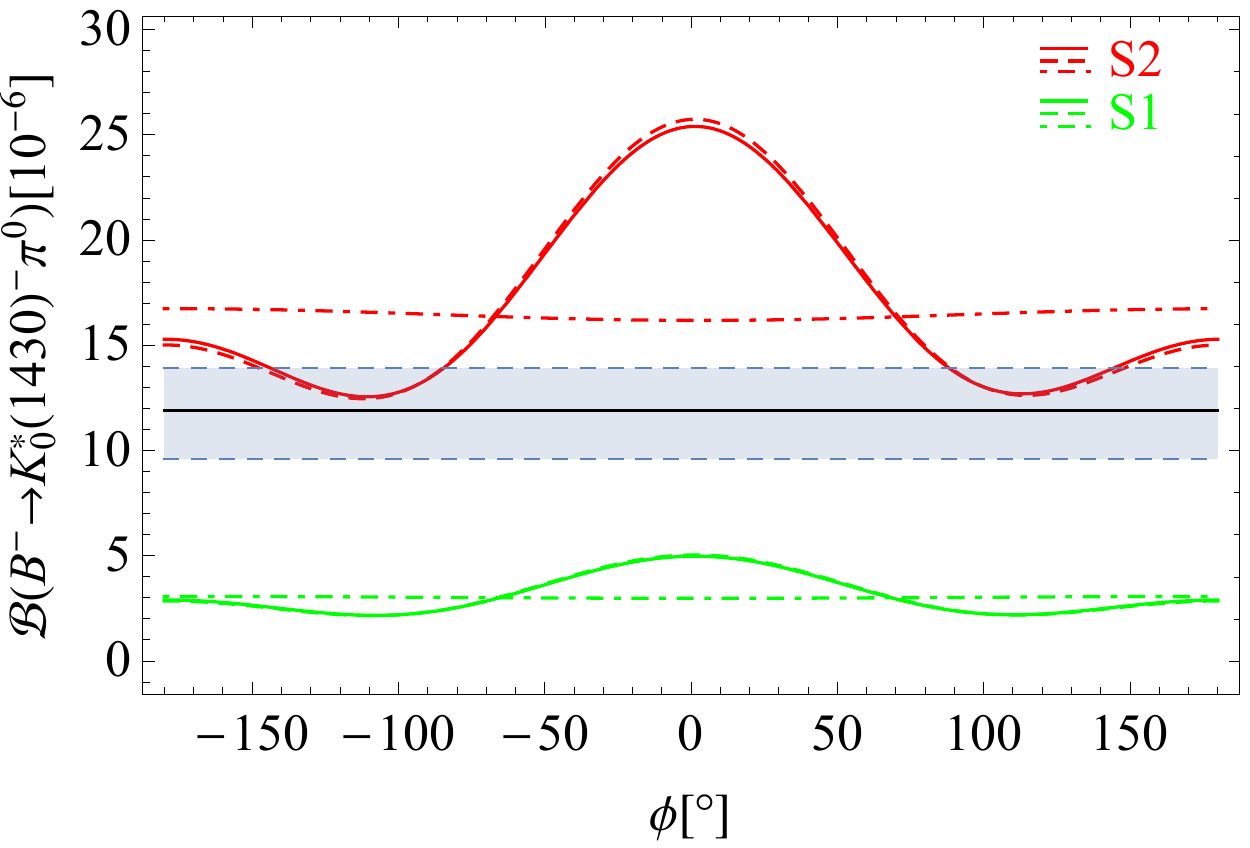}}\quad
\subfigure[]{\includegraphics[scale=0.4]{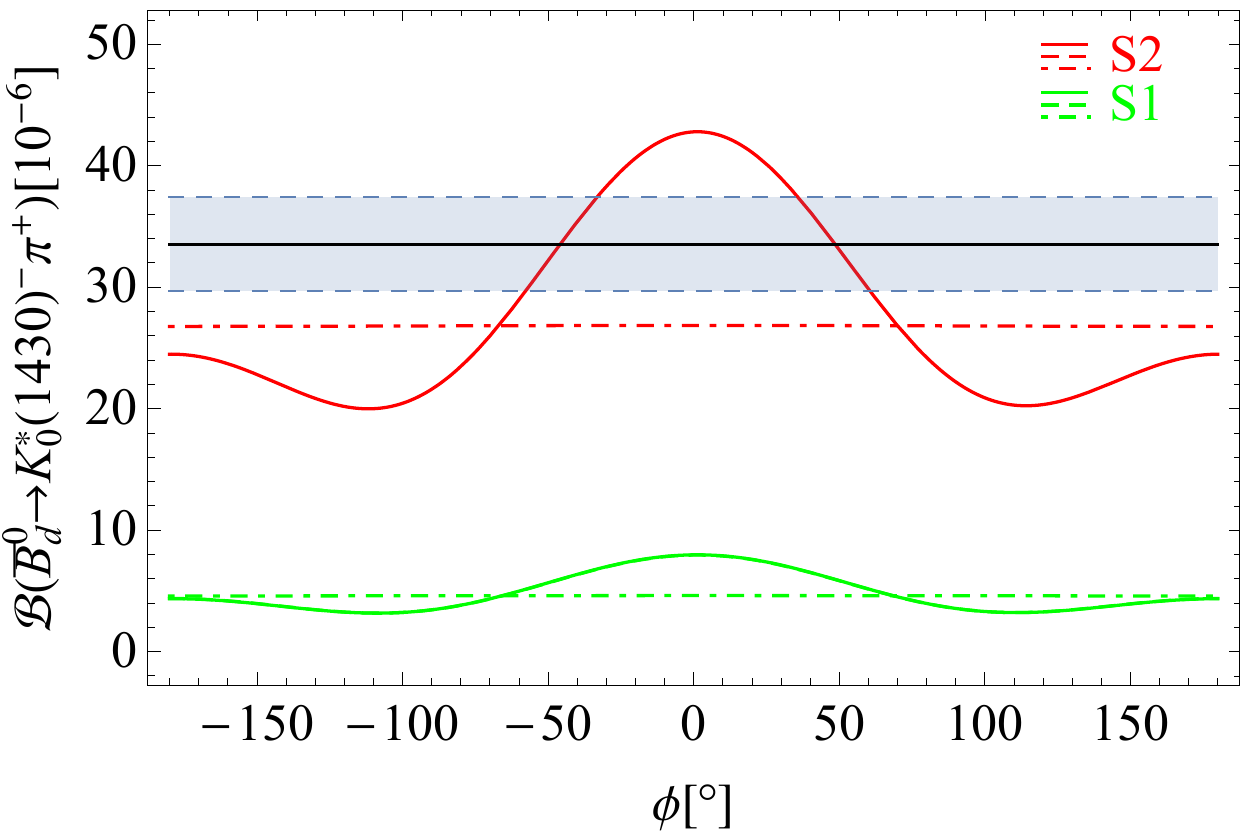}}\\
\subfigure[]{\includegraphics[scale=0.4]{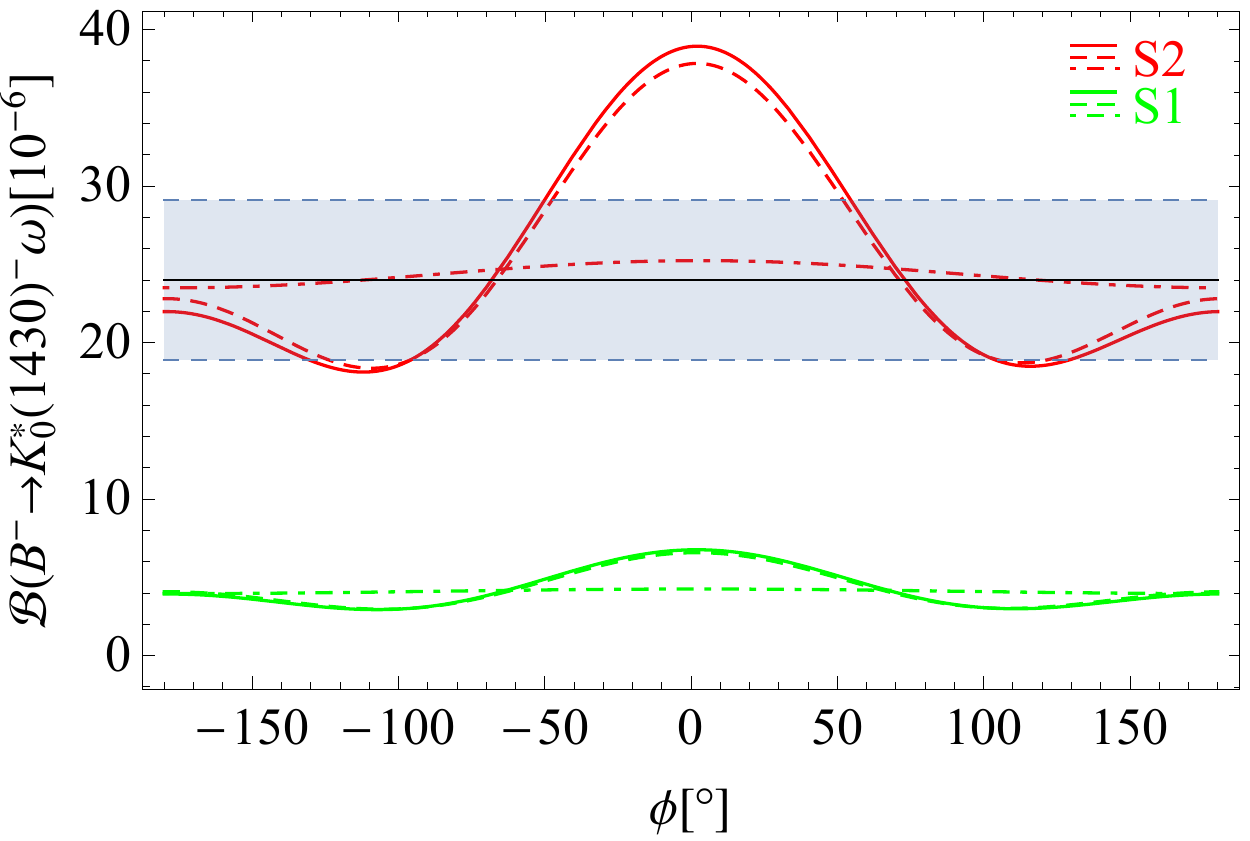}}\quad
\subfigure[]{\includegraphics[scale=0.4]{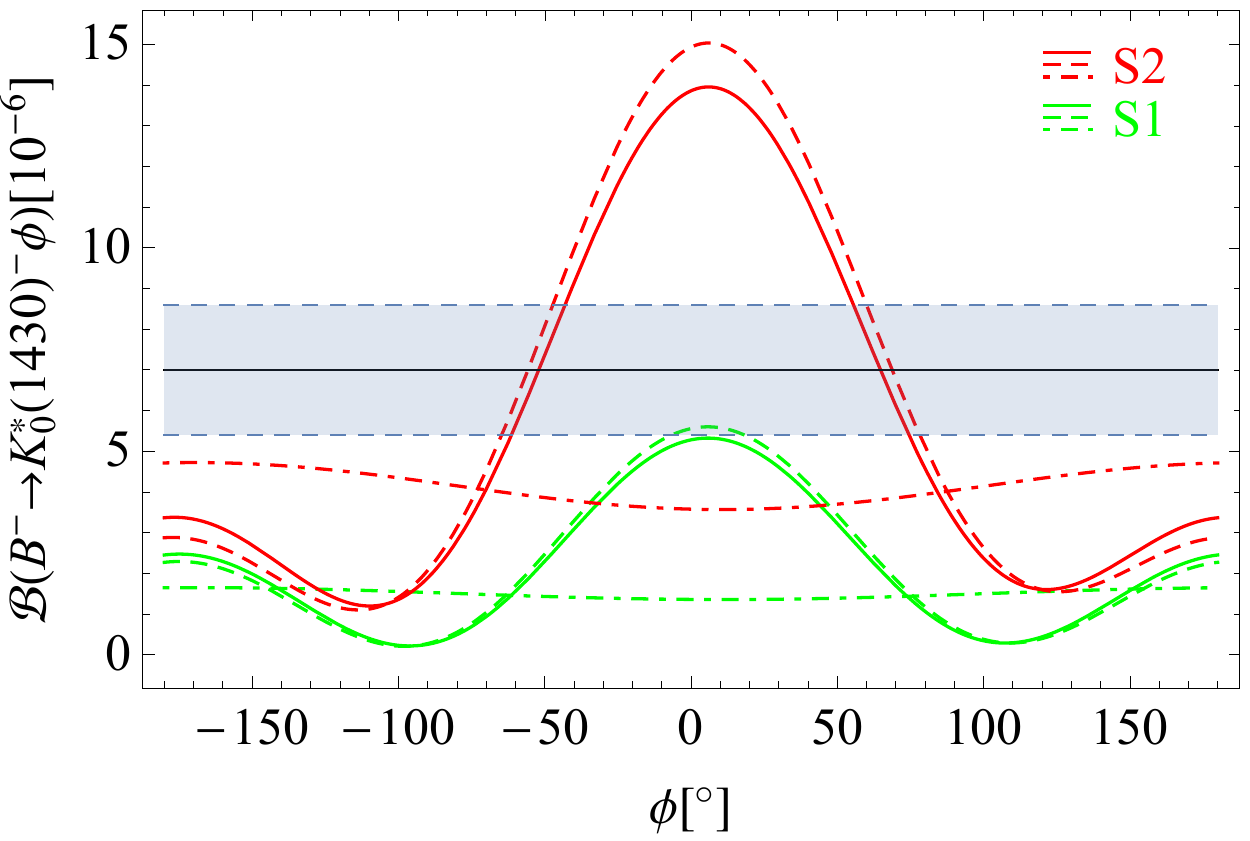}}\quad
\subfigure[]{\includegraphics[scale=0.4]{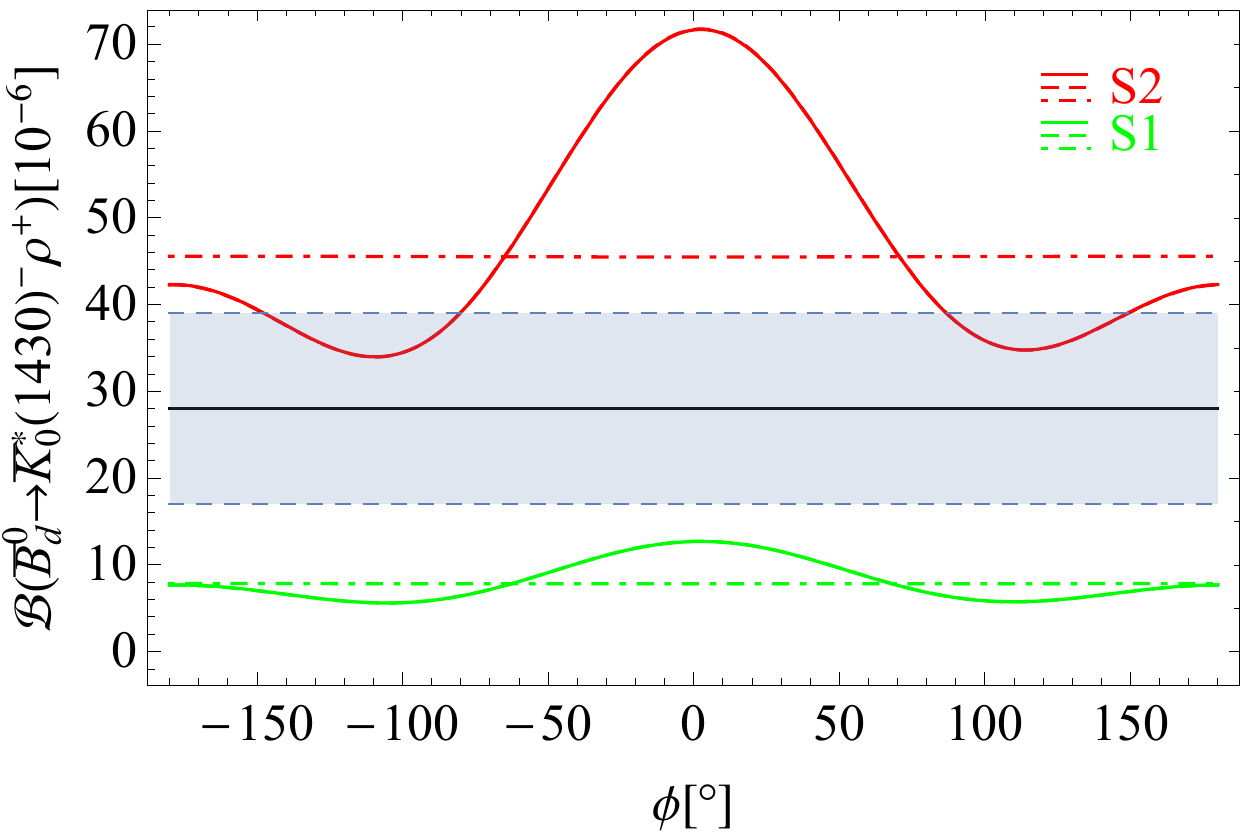}}\\
\subfigure[]{\includegraphics[scale=0.4]{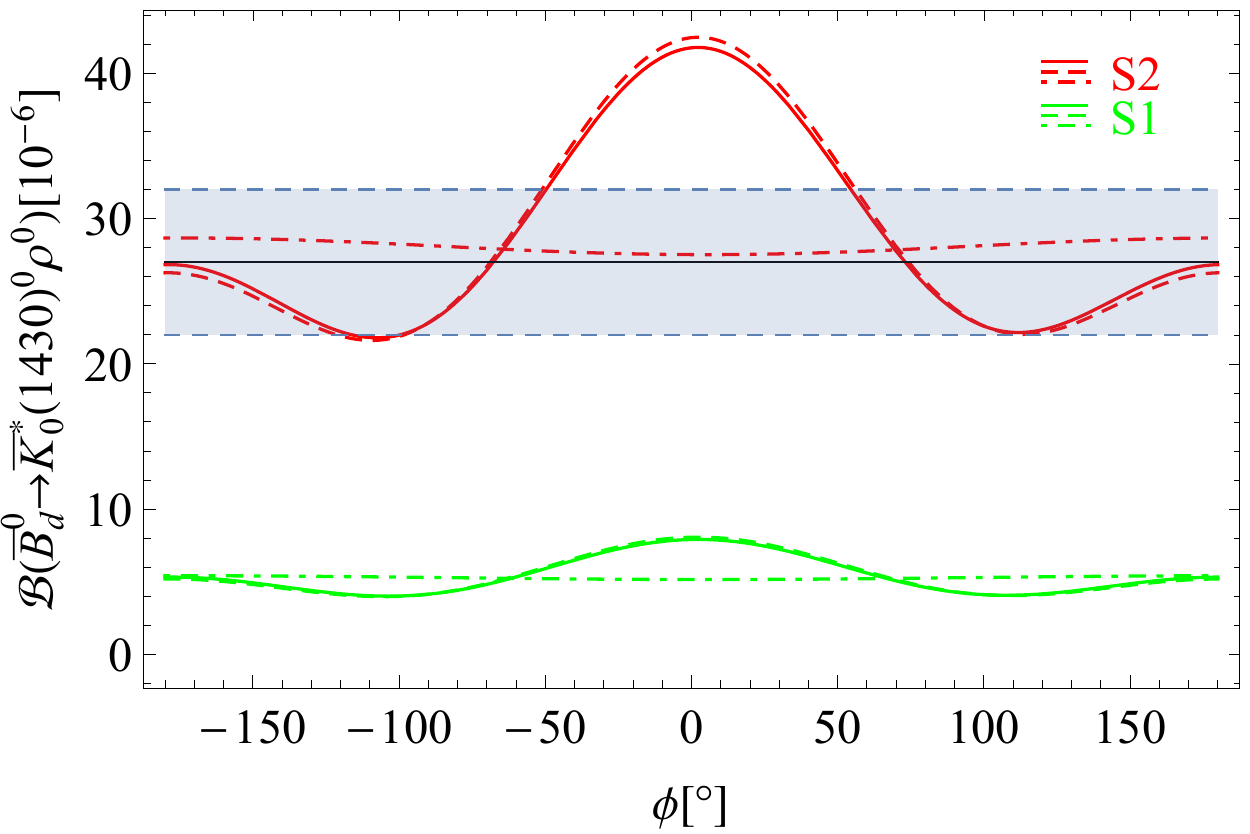}}\quad
\subfigure[]{\includegraphics[scale=0.4]{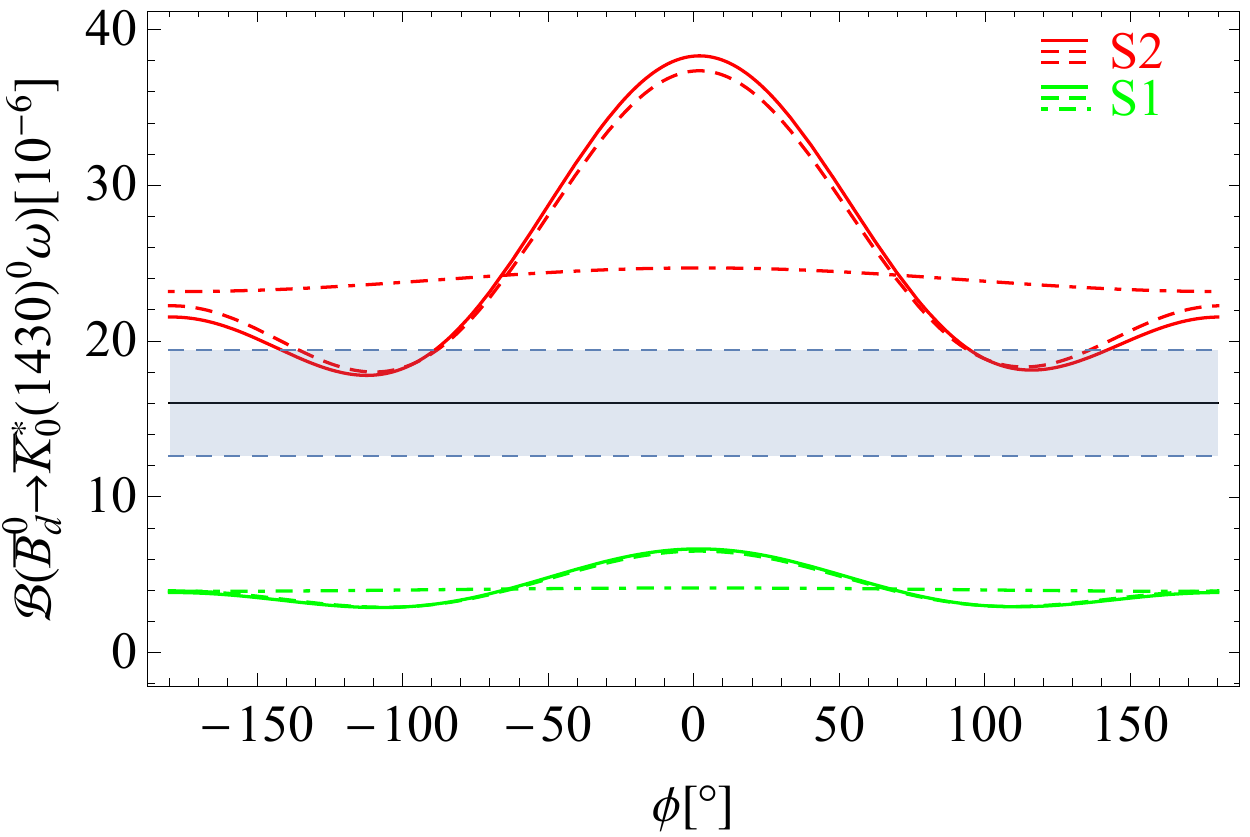}}\quad
\subfigure[]{\includegraphics[scale=0.4]{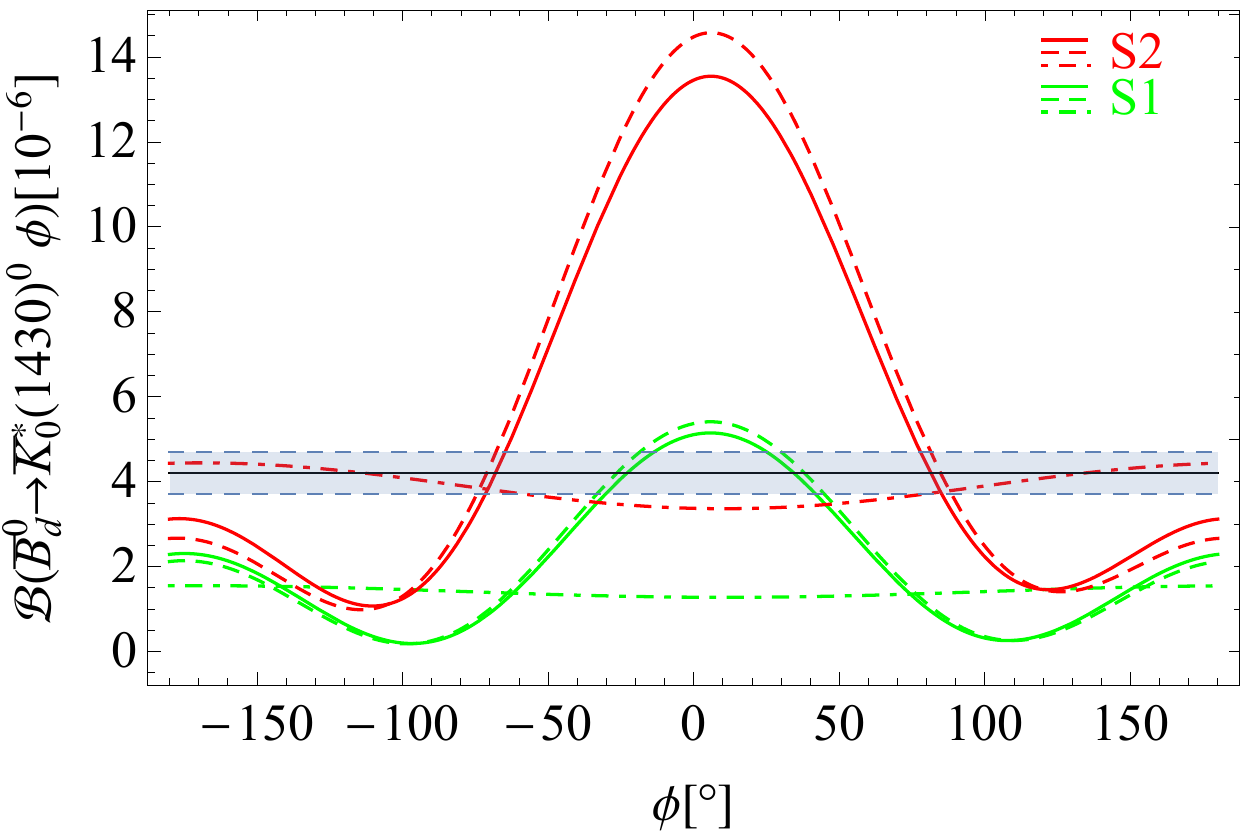}}
\caption{{The dependences of measured  ${\cal B}\left(B_{u,d,s} \to K_0^{*}(1430) (\pi\,,\rho\,,\omega\,,\phi)\right)$ on $\phi_A$ with $(\rho_A,\rho_H)={(1,0)}$ and $\phi_H$ with $(\rho_A,\rho_H)={(0,1)}$ are shown by the dash and dash-dotted lines, respectively. The solid lines are plotted with the simplification that $X_H(\rho_H,\phi_H)=X_A(\rho_A,\phi_A)$ with $\rho_A=1$. The shaded region is the experimental result with $1\sigma$ error bar. See text for further discussion.}   }
 \label{fig:Bphi}
\end{center}
\end{figure}

\begin{figure}[h]
\begin{center}
\subfigure[]{\includegraphics[scale=0.55]{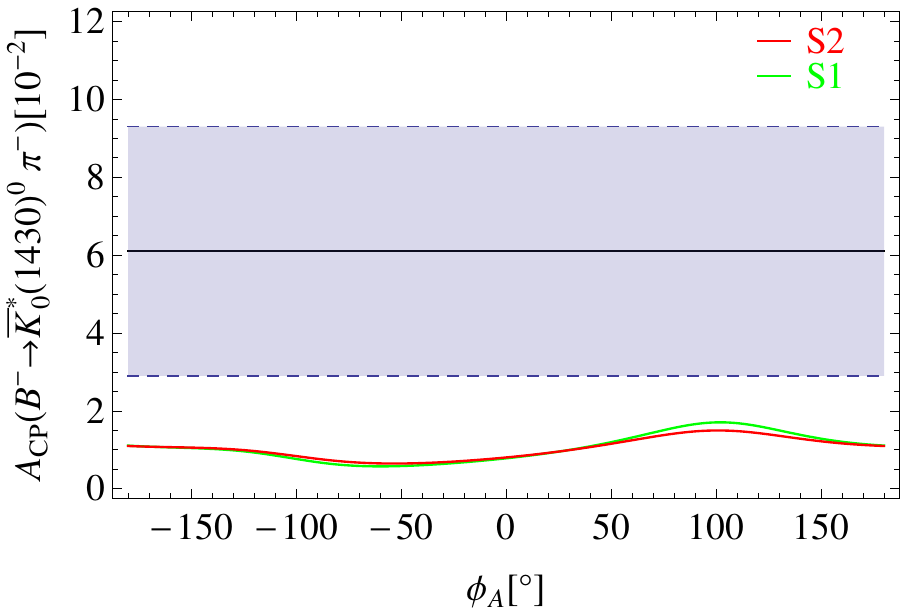}}\quad
\subfigure[]{\includegraphics[scale=0.55]{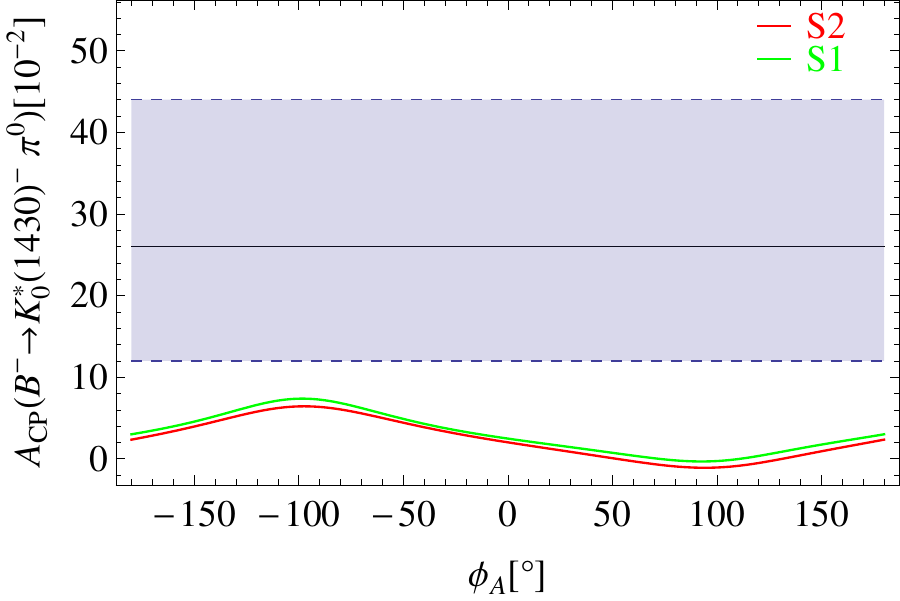}}\quad
\subfigure[]{\includegraphics[scale=0.55]{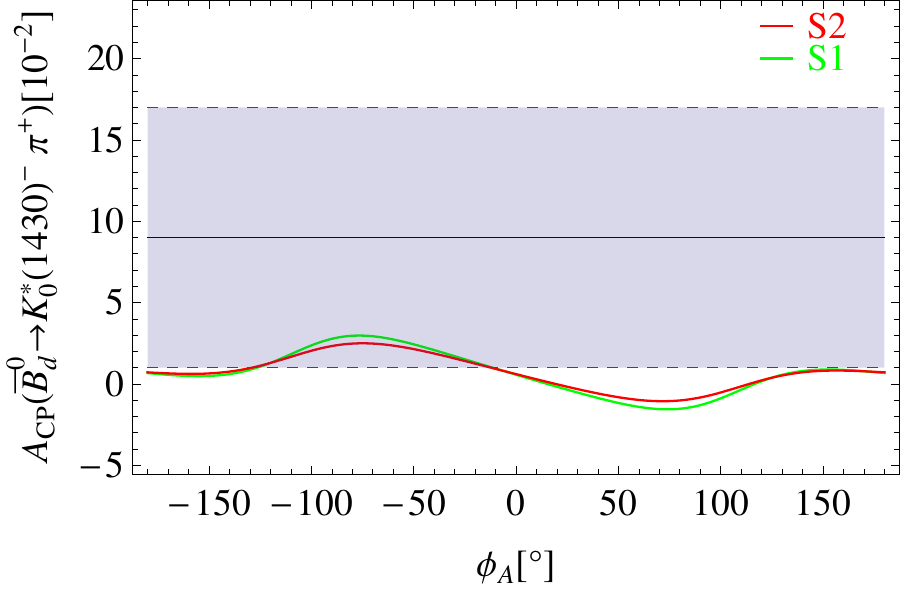}}\\
\subfigure[]{\includegraphics[scale=0.55]{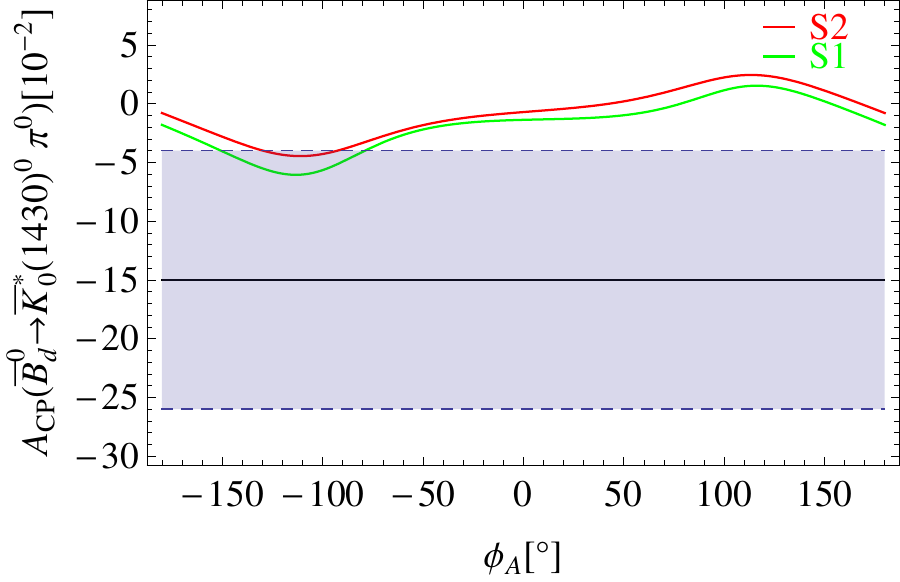}}\quad
\subfigure[]{\includegraphics[scale=0.55]{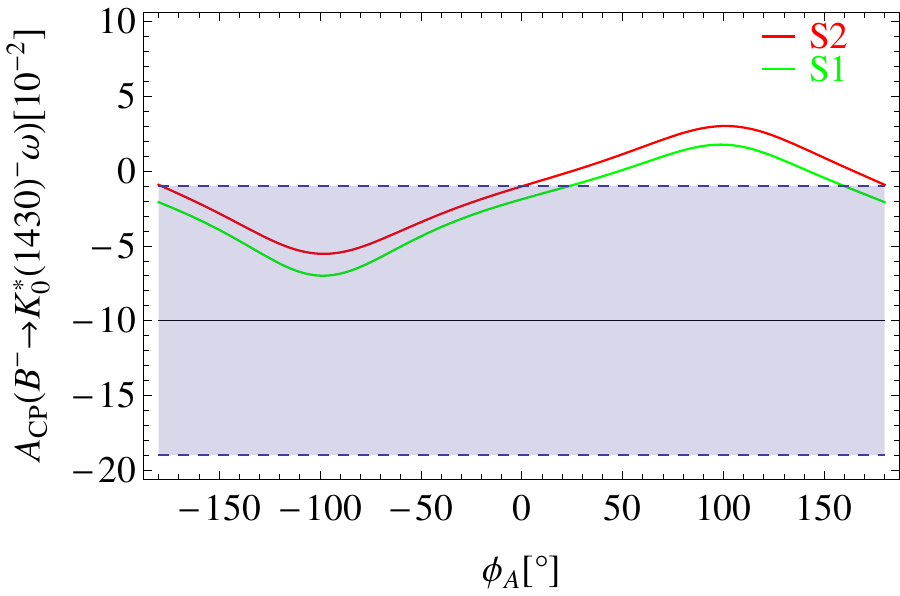}}\quad
\subfigure[]{\includegraphics[scale=0.55]{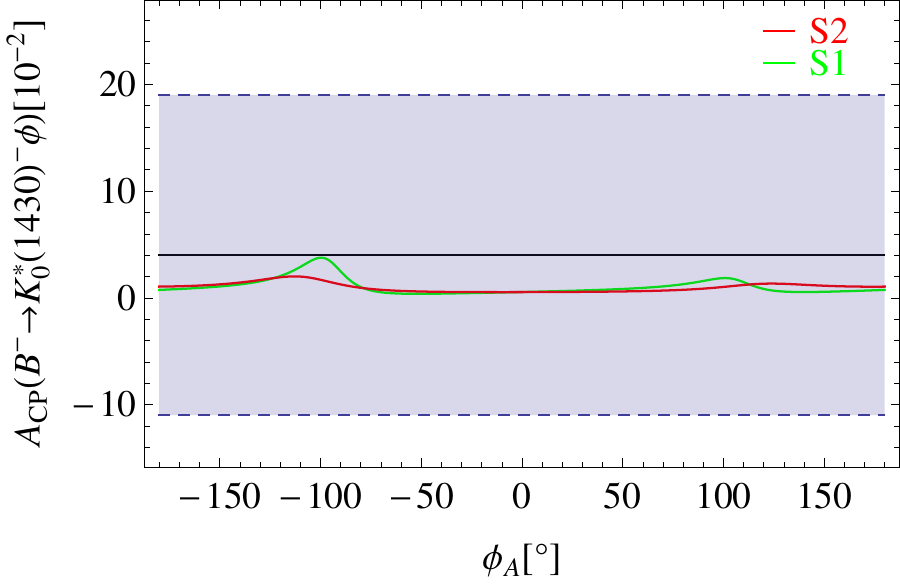}}\\
\subfigure[]{\includegraphics[scale=0.55]{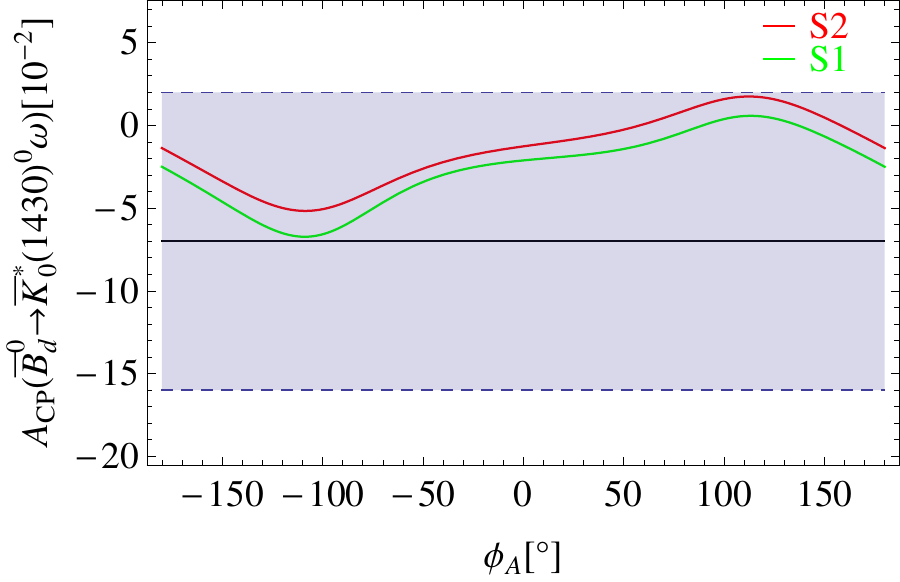}}\quad
\subfigure[]{\includegraphics[scale=0.55]{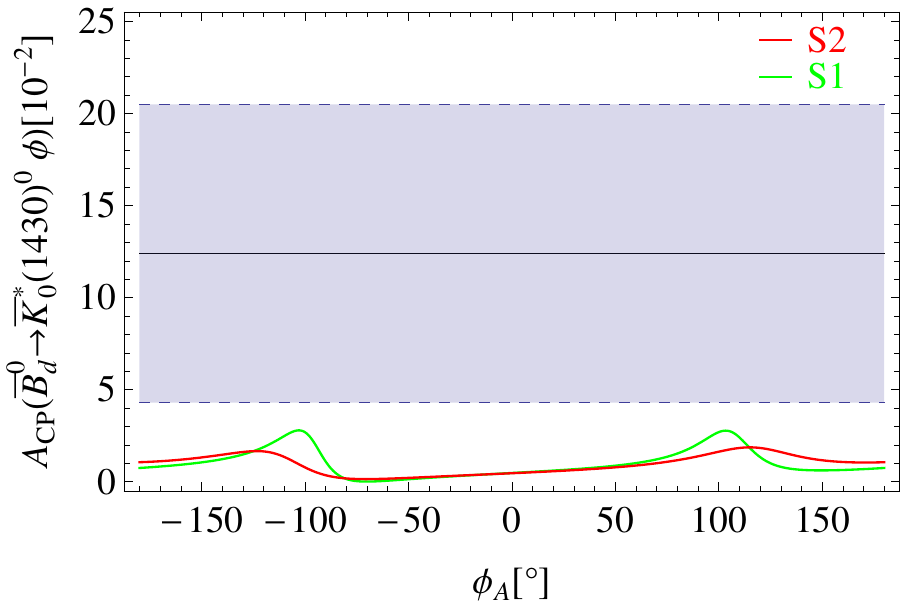}}
\caption{The dependences of measured  $A_{CP}\left(B_{u,d} \to K_0^{*}(1430) (\pi\,,\rho\,,\omega\,,\phi)\right)$ on $\phi_A$ with  $\rho_A=1$ in S1 and S2. The shaded region is the experimental result with $1\sigma$ error bar. See text for further discussion.   }
 \label{fig:CPphi}
\end{center}
\end{figure}

 Using the inputs given above, we then  present our numerical results and discussion. The QCDF approach itself cannot give the exact values of the end-point parameters {$X_H(\rho_H,\phi_H)$ in the hard-spectator scattering corrections and} $X_A(\rho_A,\phi_A)$ in the annihilation amplitudes, which can only be determined by fitting to the experimental data. The current data indicate that {$\rho_{A(H)}$} is at the level of 1, but {$\phi_{A(H)} $} is generally different for kinds of decay types.  For instance,  $\phi_A=-55^\circ$, $-22^\circ$ and $-70^\circ$ with $\rho_A=1$~(the so-called ``Scenario S4'' in Ref.~\cite{Beneke:2003zv}) are favored by the $B\to PP$, $PV$ and  $VP$ decays, respectively~\cite{Beneke:2003zv,Cheng:2009c1,Cheng:2009c2}. In order to clearly show the effects of end-point contributions, {we plot the dependences of  the measured ${\cal B}(B_{u,d} \to K_0^{*}(1430) (\pi\,,\rho\,,\omega\,,\phi))$ on the end-point parameters in Fig.~\ref{fig:Bphi}, in addition, the experimental data are also shown in such figure for comparison. The dash and dash-dotted lines show the dependences on  $\phi_A$ with $(\rho_A,\rho_H)={(1,0)}$ and $\phi_H$ with $(\rho_A,\rho_H)={(0,1)}$, respectively. Comparing the dash with the dash-dotted lines, it can be found that the measured decay modes are very sensitive to $X_A(\rho_A,\phi_A)$, but the effect of $X_H(\rho_H,\phi_H)$ is trivial. It can be easily understood because these decay modes are penguin-dominated, and the hard-spectator scattering corrections may be significant only when the decays are dominated by the color-suppressed tree contribution~\cite{Beneke:2003zv}. Therefore, we assume that the end-point parameters are universal in the annihilation and  hard-spectator scattering corrections, {\it i.e.}, $X_H(\rho_H,\phi_H)=X_A(\rho_A,\phi_A)$, in our following discussions for simplification.}~\footnote{{In the decay modes considered in this work, $\bar{B}_s\to K_0^{*0} (\pi^0,\rho^0,\omega)$ decays are color-suppressed tree dominated, thus the simplification  $X_H(\rho_H,\phi_H)=X_A(\rho_A,\phi_A)$ may affect our following predictions for these decays.  } } Under this assumption,  the dependences of  the measured branching fractions and direct CP asymmetries of $B_{u,d} \to K_0^{*}(1430) (\pi\,,\rho\,,\omega\,,\phi)$ decays on the end-point parameters are shown in Figs.~\ref{fig:Bphi} and \ref{fig:CPphi}~(solid lines), respectively.
\begin{itemize}
\item From Figs.~\ref{fig:Bphi} and \ref{fig:CPphi}, it can be found that the scenarios S1 and S2 give significantly different results for the branching fractions, but the direct CP asymmetries are similar with each other. As shown by Fig.~\ref{fig:Bphi}, the results based on S2 could agree well with the experimental data at the level of $1\sigma$ when a proper value of $\phi_A$ is taken; however,  the results for most of  decay modes  based on S1 are much smaller than data for any value of  $\phi_A$. This implies that the measured $B_{u,d} \to K_0^{*}(1430) P$ and $K_0^{*}(1430) V$ decays favor  $K_0^{*}(1430)$ as the lowest-lying $(s,u/d)$ state rather than it as the first excited one.  {In addition, this conclusion would not change even if the $20\%$ uncertainties of form factors caused by different approaches are considered.} Therefore, our following discussion is mainly based on S2, unless otherwise stated.
\item In S2, it can be found from Figs.~\ref{fig:Bphi} that $\phi_A\sim [-50^\circ,-100^\circ]$ and $[50^\circ,100^\circ]$ is allowed by the branching fractions, while  a negative  $\phi_A$ is favored by the CP asymmetries as  shown by Fig.~\ref{fig:CPphi}. In our following numerical calculation, we take
\begin{align}
(\rho_A,\phi_A)=(1,-55^\circ)  \quad\text{and}\quad (1,-64^\circ)
\end{align}
for $B\to SP$ and $SV$ decay modes, respectively.
\end{itemize}

\begin{table}[t]
\caption{Branching fractions (in units of ${10}^{-6}$) of $ B_{u,d,s}\to K_0^{*}(1430)P\,(P=\pi,K) $ decays. }
\vspace{-0.7cm}\footnotesize
\begin{center}\setlength{\tabcolsep}{5pt}
\begin{tabular}{lcccccccc}
\hline\hline
Decay modes &S2 &S1 &QCDF(S2) &pQCD &data\\ \hline
$B^- \to  K_0^{*}(1430)^- K^0 $ &$0.08^{+0.00+0.01+0.03}_{-0.00-0.02-0.03}$ &$0.06^{+0.00+0.01+0.04}_{-0.00-0.01-0.03}$
&$0.11^{+0.05}_{-0.04}$~\cite{Li:2015zra} &$0.38\pm0.22$~\cite{Wang:2020wcm} & \\
$B^- \to K_0^{*}(1430)^0 K^- $  &$3.19^{+0.18+0.36+1.73}_{-0.14-0.62-1.39}$ &$0.54^{+0.03+0.06+0.35}_{-0.02-0.11-0.27}$
&$3.37^{+1.03}_{-0.85}$~\cite{Li:2015zra} &$3.99\pm1.38$~\cite{Wang:2020wcm} &$0.38\pm0.13$  \\
$B^- \to \bar K_0^{*}(1430)^0 \pi^- $ &$33.37^{+1.48+4.28+18.52}_{-1.04-7.29-14.95}$ &$5.85^{+0.26+0.75+3.88}_{-0.18-1.27-2.99}$
&$12.9^{+4.6}_{-3.7}$~\cite{Cheng:2013fba} &$47.6^{+11.3}_{-10.1}$~\cite{Shen:2007swz} &$39^{+6}_{-5}$ \\
&&&&$36.6\pm11.3$~\cite{Wang:2020wcm} & \\
$B^- \to  K_0^{*}(1430)^- \pi^0 $& $18.42^{+0.82+2.18+9.52}_{-0.58-3.73-7.80}$ &$3.41^{+0.15+0.39+2.03}_{-0.11-0.67-1.61}$
&$7.4^{+2.4}_{-1.9}$~\cite{Cheng:2013fba} &$28.8^{+6.8}_{-6.1}$~\cite{Shen:2007swz} &${11.9^{+2.0}_{-2.3}}$ \\
&&&&$12.7\pm4.2$~\cite{Wang:2020wcm}&\\
\hline
$\bar B_d^0 \to \bar K_0^{*}(1430)^0 K^0 $ &$0.08^{+0.00+0.01+0.04}_{-0.00-0.02-0.03}$
&$0.06^{+0.00+0.01+0.04}_{-0.00-0.01-0.03}$
&$0.24^{+0.12}_{-0.09}$~\cite{Li:2015zra} &$0.49\pm0.33$~\cite{Wang:2020wcm} & \\
$\bar B_d^0 \to  K_0^{*}(1430)^0 \bar K^0 $ &$2.97^{+0.16+0.33+1.61}_{-0.13-0.57-1.29}$&
$0.50^{+0.03+0.06+0.33}_{-0.02-0.10-0.25}$
&$4.05^{+1.34}_{-1.08}$~\cite{Li:2015zra} &$4.61\pm1.50$~\cite{Wang:2020wcm} & \\
$\bar B_d^0 \to  K_0^{*}(1430)^- K^+ $ &$0.08^{+0.01+0.02+0.04}_{-0.01-0.02-0.03}$
&$0.02^{+0.00+0.00+0.01}_{-0.00-0.00-0.01}$
&$0.11^{+0.07}_{-0.05}$~\cite{Li:2015zra} &$0.09\pm0.06$~\cite{Wang:2020wcm} & \\
$\bar B_d^0 \to  K_0^{*}(1430)^+ K^- $ &$0.09^{+0.01+0.02+0.05}_{-0.01-0.03-0.04}$
&$0.05^{+0.00+0.01+0.04}_{-0.00-0.01-0.03}$
&$0.06^{+0.05}_{-0.03}$~\cite{Li:2015zra} &$0.62\pm0.40$~\cite{Wang:2020wcm} & \\
$\bar B_d^0 \to K_0^{*}(1430)^- \pi^+ $ &$30.55^{+1.36+3.85+16.97}_{-0.97-6.56-13.70}$
&$5.31^{+0.24+0.66+3.52}_{-0.17-1.13-2.72}$
&$13.8^{+4.5}_{-3.6}$~\cite{Cheng:2013fba} &$43.0^{+10.2}_{-9.1}$~\cite{Shen:2007swz} &$33.5^{+3.9}_{-3.8}$\\    &&&&$33.4\pm10.2$~\cite{Wang:2020wcm}\\
$\bar B_d^0 \to  \bar K_0^{*}(1430)^0  \pi^0 $ &$13.78^{+0.61+1.89+8.24}_{-0.44-3.20-6.54}$
&$2.26^{+0.10+0.32+1.68}_{-0.07-0.54-1.25}$
&$5.6^{+2.6}_{-1.3}$~\cite{Cheng:2013fba} &$18.4^{+4.4}_{-3.9}$~\cite{Shen:2007swz} & \\
&&&&$22.4\pm6.6$~\cite{Wang:2020wcm}\\
\hline
$\bar B_s^0 \to \bar K_0^{*}(1430)^0 K^0 $ &$29.70^{+1.31+3.74+16.51}_{-0.93-6.37-13.32}$
&$5.40^{+0.24+0.66+3.59}_{-0.17-1.13-2.76}$ &&& \\
$\bar B_s^0 \to  K_0^{*}(1430)^0 \bar K^0 $ &$3.08^{+0.14+0.44+1.51}_{-0.10-0.72-1.17}$
&$1.22^{+0.05+0.22+0.78}_{-0.04-0.35-0.59}$ &&& \\
$\bar B_s^0 \to  K_0^{*}(1430)^- K^+ $ &$29.79^{+1.33+3.70+16.56}_{-0.95-6.30-13.36}$
&$5.37^{+0.24+0.65+3.57}_{-0.17-1.11-2.75}$ &&& \\
$\bar B_s^0 \to K_0^{*}(1430)^+ K^- $ &$3.94^{+0.18+0.47+1.57}_{-0.13-0.79-1.21}$ & $1.82^{+0.09+0.24+0.81}_{-0.07-0.40-0.63}$ &&&\\
$\bar B_s^0 \to K_0^{*}(1430)^+ \pi^- $ &$6.41^{+0.49+0.04+1.22}_{-0.44-0.02-1.14}$&
$4.51^{+0.35+0.01+0.98}_{-0.31-0.01-0.91}$
&&$37.0^{+14.0}_{-10.0}$~\cite{Zhang:2012z} \\
$\bar B_s^0 \to  K_0^{*}(1430)^0 \pi^0 $ &$0.70^{+0.05+0.07+0.31}_{-0.04-0.12-0.26}$&
$0.21^{+0.01+0.02+0.09}_{-0.01-0.03-0.07}$
&&$0.41^{+0.10}_{-0.07}$~\cite{Zhang:2012z}\\
\hline\hline
\end{tabular}
\end{center}
\label{tab:BSP}
\end{table}
\begin{table}[h]
\caption{Branching fractions (in units of ${10}^{-6}$) of  $ B_{u,d,s}\to K_0^{*}(1430)V\,(V=K^*\,, \rho\,, \omega\,, \phi) $ decays. }
\vspace{-0.7cm}\footnotesize
\begin{center}\setlength{\tabcolsep}{5pt}
\begin{tabular}{lccccc}
\hline\hline
Decay modes  &S2 &S1 &QCDF(S2) &pQCD(S2) &data\\ \hline
$B^- \to  K_0^{*}(1430)^- K^{*0} $ &$0.24^{+0.01+0.01+0.05}_{-0.01-0.01-0.04}$ &$0.07^{+0.00+0.00+0.01}_{-0.00-0.00-0.01}$
&$0.01^{+0.01}_{-0.01}$~\cite{Li:2015zra} &$1.3^{+0.5}_{-0.3}$~\cite{Liu:2013lxz} & \\
$B^- \to K_0^{*}(1430)^0 K^{*-} $ &$2.48^{+0.14+0.30+1.38}_{-0.11-0.52-1.12}$
&$0.43^{+0.02+0.05+0.29}_{-0.02-0.09-0.22}$
&$2.17^{+0.55}_{-0.47}$~\cite{Li:2015zra} &$1.5^{+0.5}_{-0.3}$~\cite{Liu:2013lxz} & \\
$B^- \to \bar K_0^{*}(1430)^0 \rho^- $ &$51.81^{+2.29+5.64+28.39}_{-1.62-9.71-22.86}$ &$8.66^{+0.38+0.97+5.71}_{-0.27-1.66-4.40}$
&$39.0^{+34.5}_{-35.8}$~\cite{Cheng:2013fba} &$12.1^{+2.8}_{-0.0}$~\cite{Zhang:2010zbud} & \\
$B^- \to  K_0^{*}(1430)^- \rho^0 $ &$20.99^{+0.93+2.55+12.72}_{-0.66-4.35-10.02}$
&$3.15^{+0.14+0.41+2.44}_{-0.10-0.70-1.80}$
&$14.8^{+3.7}_{-3.2}$~\cite{Cheng:2013fba} &$8.4^{+2.3}_{-0.0}$~\cite{Zhang:2010zbud} & \\
$B^- \to  K_0^{*}(1430)^- \omega $ &$25.08^{+1.11+2.82+13.98}_{-0.78-4.84-11.20}$
&$4.11^{+0.18+0.48+2.79}_{-0.13-0.82-2.13}$
&$21.5^{+5.8}_{-4.9}$~\cite{Cheng:2013fba} &$7.4^{+2.1}_{-1.5}$~\cite{Zhang:2010zbud} &$24.0\pm5.1$ \\
$B^- \to  K_0^{*}(1430)^- \phi $ &$4.97^{+0.22+0.21+1.08}_{-0.16-0.37-0.96}$ &$1.29^{+0.06+0.08+0.35}_{-0.04-0.11-0.22} $
&$3.80^{+0.7}_{-0.6}$~\cite{Cheng:2013fba} &$25.6^{+6.2}_{-5.4}$~\cite{Kim:2010cyw} &$7.0\pm1.6$ \\
\hline
$\bar B_d^0 \to \bar K_0^{*}(1430)^0 K^{*0} $ &$0.22^{+0.01+0.01+0.05}_{-0.01-0.01-0.04}$
&$0.06^{+0.00+0.00+0.01}_{-0.00-0.00-0.01}$
&&& $<3.3$ \\
$\bar B_d^0 \to  K_0^{*}(1430)^0 \bar K^{*0} $ &$2.30^{+0.13+0.28+1.28}_{-0.10-0.48-1.03}$
&$0.39^{+0.02+0.05+0.26}_{-0.02-0.08-0.20}$
&$2.01^{+0.54}_{-0.45}$~\cite{Li:2015zra} && \\
$\bar B_d^0 \to  K_0^{*}(1430)^- K^{*+} $ &$0.03^{+0.00+0.00+0.02}_{-0.00-0.00-0.01}$
&$0.01^{+0.00+0.00+0.01}_{-0.00-0.00-0.01}$
&$0.18^{+0.13}_{-0.08}$~\cite{Li:2015zra} && \\
$\bar B_d^0 \to  K_0^{*}(1430)^+ K^{*-} $ &$0.04^{+0.00+0.00+0.03}_{-0.00-0.00-0.02}$
&$0.05^{+0.00+0.00+0.04}_{-0.00-0.00-0.03}$
&$0.01^{+0.01}_{-0.01}$~\cite{Li:2015zra} && \\
$\bar B_d^0 \to K_0^{*}(1430)^- \rho^+ $ &$45.92^{+2.03+5.05+25.14}_{-1.43-8.69-20.24}$
&$7.67^{+0.34+0.86+5.05}_{-0.24-1.48-3.89} $
&$36.3^{+8.5}_{-7.4}$~\cite{Cheng:2013fba} &$10.5^{+2.7}_{-0.0}$~\cite{Zhang:2010zbud} &$28.0\pm11.0$ \\
$\bar B_d^0 \to  \bar K_0^{*}(1430)^0  \rho^0 $ &$27.95^{+1.24+2.78+13.95}_{-0.87-4.82-11.45}$ &$5.10^{+0.23+0.50+2.91}_{-0.16-0.86-2.32} $
&$23.4^{+5.1}_{-4.5}$~\cite{Cheng:2013fba} &$4.8^{+1.1}_{-0.0}$~\cite{Zhang:2010zbud} &$27.0\pm 5.0$ \\
$\bar B_d^0 \to  \bar K_0^{*}(1430)^0  \omega $ &$24.71^{+1.10+2.74+13.76}_{-0.78-4.71-11.03}$
&$4.06^{+0.18+0.46+2.75}_{-0.13-0.80-2.10}$
&$21.9^{+5.9}_{-5.0}$~\cite{Cheng:2013fba} &$9.3^{+2.1}_{-2.0}$~\cite{Zhang:2010zbud} &$16.0\pm3.4$  \\
$\bar B_d^0 \to  \bar K_0^{*}(1430)^0  \phi $ &$4.80^{+0.21+0.21+1.46}_{-0.15-0.38-1.18}$
&$1.25^{+0.06+0.08+0.42}_{-0.04-0.11-0.29} $
&$3.7^{+0.8}_{-0.6}$~\cite{Cheng:2013fba} &$23.6^{+5.6}_{-5.0}$~\cite{Kim:2010cyw} &$4.2\pm0.5$\\
\hline
$\bar B_s^0 \to \bar K_0^{*}(1430)^0 K^{*0} $ &$29.87^{+1.32+3.28+17.95}_{-0.93-5.64-14.51}$
&$5.08^{+0.22+0.56+3.57}_{-0.16-0.97-2.77}$ &&& \\
$\bar B_s^0 \to  K_0^{*}(1430)^0 \bar K^{*0} $ &$5.11^{+0.23+0.29+1.43}_{-0.16-0.50-1.17}$
&$2.15^{+0.09+0.07+0.50}_{-0.07-0.12-0.42}$ &&& \\
$\bar B_s^0 \to  K_0^{*}(1430)^- K^{*+} $ &$28.39^{+1.26+3.15+17.04}_{-0.89-5.42-13.77}$
&$4.83^{+0.21+0.54+3.39}_{-0.15-0.93-2.63}$ &&& \\
$\bar B_s^0 \to K_0^{*}(1430)^+ K^{*-} $ &$5.42^{+0.28+0.32+1.63}_{-0.21-0.56-1.34}$
&$2.17^{+0.13+0.08+0.51}_{-0.11-0.13-0.42}$ &&& \\
$\bar B_s^0 \to K_0^{*}(1430)^+ \rho^- $ &$17.77^{+1.36+0.12+2.74}_{-1.22-0.07-2.56}$
&$12.57^{+0.97+0.05+2.30}_{-0.86-0.03-2.12}$
&&$108^{+25}_{-23}$~\cite{Zhang:2010zbs} & \\
$\bar B_s^0 \to  K_0^{*}(1430)^0  \rho^0 $ &$1.28^{+0.10+0.11+0.56}_{-0.09-0.20-0.46}$
&$0.35^{+0.03+0.02+0.16}_{-0.02-0.04-0.13} $
&&$0.96^{+0.22}_{-0.20}$~\cite{Zhang:2010zbs} & \\
$\bar B_s^0 \to  K_0^{*}(1430)^0  \omega $ &$1.45^{+0.11+0.12+0.57}_{-0.10-0.20-0.48}$
&$0.44^{+0.03+0.03+0.17}_{-0.03-0.05-0.14} $
&&$0.86^{+0.21}_{-0.18}$~\cite{Zhang:2010zbs} & \\
$\bar B_s^0 \to  K_0^{*}(1430)^0  \phi $ &$1.90^{+0.11+0.22+1.03}_{-0.08-0.38-0.84}$
&$0.35^{+0.02+0.04+0.21}_{-0.01-0.07-0.17} $
&&$0.95^{+0.25}_{-0.17}$~\cite{Zhang:2010zbs} & \\
\hline\hline
\end{tabular}
\end{center}
\label{tab:BSV}
\end{table}
\begin{table}[t]
\caption{The direct CP asymmetries (in units of $\%$) of $ B_{u,d,s}\to K_0^{*}(1430)P\,(P=\pi,K) $decays.}
\vspace{-0.7cm}\footnotesize
\begin{center}\setlength{\tabcolsep}{5pt}
\begin{tabular}{lccccc}
\hline\hline
Decay modes  &S2 &S1 &QCDF(S2) &pQCD(S2) &data\\ \hline
$B^- \to  K_0^{*}(1430)^- K^0 $ &$30.32^{+1.08+0.73+1.23}_{-1.11-0.42-1.43}$ &$18.58^{+0.71+1.79+2.67}_{-0.72-0.79-1.94}$
&$-22.51^{+4.90}_{-7.57}$~\cite{Li:2015zra} && \\
$B^- \to K_0^{*}(1430)^0 K^- $ &$-16.56^{+0.64+0.28+0.88}_{-0.63-0.56-0.84}$ &$-12.25^{+0.48+1.42+2.23}_{-0.47-1.47-2.25} $
&$-2.60^{+1.61}_{-1.76}$~\cite{Li:2015zra} &&$10\pm17$ \\
$B^- \to \bar K_0^{*}(1430)^0 \pi^- $ &$0.65^{+0.02+0.02+0.05}_{-0.02-0.01-0.06}$
&$0.57^{+0.02+0.07+0.13}_{-0.02-0.07-0.13}$
&$1.3^{+0.1}_{-0.1}$~\cite{Cheng:2013fba} &&$6.1\pm 3.2$ \\
$B^- \to  K_0^{*}(1430)^- \pi^0 $ &$4.71^{+0.15+0.30+0.43}_{-0.15-0.14-0.34}$
&$5.20^{+0.16+0.37+0.64}_{-0.16-0.18-0.46} $
&$3.0^{+0.4}_{-0.4}$~\cite{Cheng:2013fba} &$3.5$~\cite{Shen:2007swz} &$26^{+18}_{-14}$\\
\hline
$\bar B_d^0 \to \bar K_0^{*}(1430)^0 K^0 $ &$0.68^{+0.03+3.08+1.34}_{-0.03-1.48-2.62}$ &$0.08^{+0.00+2.05+0.81}_{-0.00-0.97-1.81}$
&&& \\
$\bar B_d^0 \to  K_0^{*}(1430)^0 \bar K^0 $ &$-20.75^{+0.79+0.13+0.61}_{-0.77-0.28-0.59}$
&$-19.68^{+0.74+1.27+1.38}_{-0.72-1.35-1.29}$
&&& \\
$\bar B_d^0 \to  K_0^{*}(1430)^- K^+ $ &$-1.38^{+0.05+0.16+0.28}_{-0.05-0.38-0.28}$
&$-9.01^{+0.32+0.80+4.82}_{-0.32-1.77-4.43}$
&&& \\
$\bar B_d^0 \to  K_0^{*}(1430)^+ K^- $ &$1.46^{+0.06+0.33+0.14}_{-0.06-0.15-0.16}$
&$3.56^{+0.14+0.25+0.79}_{-0.14-0.13-1.46}$
&&& \\
$\bar B_d^0 \to K_0^{*}(1430)^- \pi^+ $ &$2.25^{+0.07+0.21+0.17}_{-0.07-0.09-0.16}$ &$2.59^{+0.08+0.31+0.50}_{-0.08-0.22-0.53}$
&$0.21^{+0.06}_{-0.06}$~\cite{Cheng:2013fba} &&$9\pm 8$ \\
$\bar B_d^0 \to  \bar K_0^{*}(1430)^0  \pi^0 $ &$-2.04^{+0.06+0.07+0.38}_{-0.06-0.16-0.54}$
&$-2.42^{+0.08+0.23+0.85}_{-0.08-0.31-1.11} $
&$-1.9^{+0.4}_{-0.5}$~\cite{Cheng:2013fba} &&$-15\pm11$ \\
\hline
$\bar B_s^0 \to \bar K_0^{*}(1430)^0 K^0 $ &$0.82^{+0.03+0.01+0.04}_{-0.03-0.00-0.05}$
&$0.77^{+0.02+0.06+0.09}_{-0.02-0.06-0.10}$
&&& \\
$\bar B_s^0 \to  K_0^{*}(1430)^0 \bar K^0 $ &$0.03^{+0.00+0.04+0.06}_{-0.00-0.07-0.04}$
&$0.17^{+0.01+0.06+0.15}_{-0.01-0.14-0.10}$
&&& \\
$\bar B_s^0 \to  K_0^{*}(1430)^- K^+ $ &$1.25^{+0.04+0.41+0.14}_{-0.04-0.19-0.14}$
&$1.78^{+0.06+0.50+0.77}_{-0.06-0.28-0.86}$
&&& \\
$\bar B_s^0 \to K_0^{*}(1430)^+ K^- $ &$-63.79^{+1.59+3.12+7.34}_{-1.56-6.56-4.83}$
&$-61.91^{+1.80+3.82+8.67}_{-1.83-8.20-2.54}$
&&& \\
$\bar B_s^0 \to K_0^{*}(1430)^+ \pi^- $ &$25.23^{+0.92+1.51+6.69}_{-0.91-2.83-6.57}$ &$15.14^{+0.55+0.92+6.42}_{-0.55-1.72-6.37} $
&&$21.0^{+3.4}_{-3.1}$~\cite{Zhang:2012z} & \\
$\bar B_s^0 \to  K_0^{*}(1430)^0 \pi^0 $ &$-58.32^{+1.82+3.32+3.83}_{-1.80-1.62-4.91}$
&$-71.72^{+2.14+4.17+10.77}_{-2.16-1.97-10.12}$
&&$95.5^{+1.2}_{-8.7}$~\cite{Zhang:2012z} & \\
\hline\hline
\end{tabular}
\end{center}
\label{tab:BSPCP}
\end{table}
\begin{table}[h]
\caption{The direct CP asymmetries (in units of $\%$) of  $ B_{u,d,s}\to K_0^{*}(1430)V\,(V=K^*\,, \rho\,, \omega\,, \phi) $ decays.}
\vspace{-0.7cm}\footnotesize
\begin{center}\setlength{\tabcolsep}{5pt}
\begin{tabular}{lccccc}
\hline\hline
Decay modes&S2 &S1 &QCDF(S2) &pQCD(S2) &data \\ \hline
$B^- \to  K_0^{*}(1430)^- K^{*0} $ &$-13.24^{+0.46+0.39+2.34}_{-0.45-0.77-2.36}$
&$-11.37^{+0.39+1.24+6.04}_{-0.38-1.46-5.30}$
&$-31.02^{+4.67}_{-6.56}$~\cite{Li:2015zra} &$-34.9^{+5.0}_{-4.5}$~\cite{Liu:2013lxz} & \\
$B^- \to K_0^{*}(1430)^0 K^{*-} $ &$-19.99^{+0.75+0.19+0.66}_{-0.73-0.10-0.66}$ &$-26.27^{+0.94+1.19+3.06}_{-0.92-1.28-3.02}$
&$0.64^{+2.54}_{-2.64}$~\cite{Li:2015zra} &$-67.9^{+4.9}_{-5.2}$~\cite{Liu:2013lxz} & \\
$B^- \to \bar K_0^{*}(1430)^0 \rho^- $ &$1.02^{+0.03+0.01+0.03}_{-0.03-0.00-0.03}$
&$1.13^{+0.04+0.06+0.17}_{-0.04-0.06-0.16}$
&$0.32^{+0.34}_{-0.29}$~\cite{Cheng:2013fba} &$-7.1^{+0.0}_{-0.0}$~\cite{Zhang:2010zbud} & \\
$B^- \to  K_0^{*}(1430)^- \rho^0 $ &$-4.64^{+0.14+0.18+0.65}_{-0.14-0.39-0.93}$
&$-6.10^{+0.19+0.30+1.09}_{-0.19-0.64-1.88}$
&$1.6^{+0.6}_{-0.6}$~\cite{Cheng:2013fba} &$6.3^{+0.0}_{-0.1}$~\cite{Zhang:2010zbud} & \\
$B^- \to  K_0^{*}(1430)^- \omega $ &$-4.25^{+0.13+0.14+0.53}_{-0.13-0.32-0.73}$
&$-5.34^{+0.17+0.23+0.83}_{-0.17-0.47-1.32}$
&$0.55^{+0.35}_{-0.34}$~\cite{Cheng:2013fba} &$6.2^{+0.0}_{-0.0}$~\cite{Zhang:2010zbud} &$-10\pm9$  \\
$B^- \to  K_0^{*}(1430)^- \phi $ &$0.68^{+0.02+0.05+0.17}_{-0.02-0.02-0.16}$
&$0.42^{+0.01+0.07+0.46}_{-0.01-0.06-0.47}$
&$0.64^{+0.02}_{-0.03}$~\cite{Cheng:2013fba} &$1.9$~\cite{Kim:2010cyw} &$4\pm15$ \\
\hline
$\bar B_d^0 \to \bar K_0^{*}(1430)^0 K^{*0} $ &$-3.95^{+0.14+0.50+2.61}_{-0.14-1.01-3.21}$
&$-4.49^{+0.16+1.78+5.25}_{-0.16-2.04-5.59}$
&&& \\
$\bar B_d^0 \to  K_0^{*}(1430)^0 \bar K^{*0} $&$-15.78^{+0.60+0.38+0.64}_{-0.59-0.17-0.61}$
&$-15.61^{+0.59+1.40+1.07}_{-0.58-1.36-1.05}$
&&& \\
$\bar B_d^0 \to  K_0^{*}(1430)^- K^{*+} $ &$-5.10^{+0.19+0.19+0.35}_{-0.20-0.38-0.39}$
&$-1.37^{+0.06+0.64+1.74}_{-0.06-0.30-2.18}$
&&$-83.9\pm0.7$~\cite{Liu:2013lxz} & \\
$\bar B_d^0 \to  K_0^{*}(1430)^+ K^{*-} $ &$-4.27^{+0.16+0.14+0.43}_{-0.16-0.29-0.52}$
&$-3.64^{+0.14+0.02+0.61}_{-0.15-0.03-0.68}$
&&$38.5^{+1.1}_{-0.8}$~\cite{Liu:2013lxz} & \\
$\bar B_d^0 \to K_0^{*}(1430)^- \rho^+ $ &$-0.15^{+0.00+0.05+0.15}_{-0.00-0.11-0.16}$
&$-0.30^{+0.01+0.08+0.27}_{-0.01-0.15-0.30}$
&$1.1^{+0.0}_{-0.0}$~\cite{Cheng:2013fba} &$-4.8^{+0.3}_{-0.0}$~\cite{Zhang:2010zbud} & \\
$\bar B_d^0 \to  \bar K_0^{*}(1430)^0  \rho^0 $ &$4.85^{+0.15+0.13+0.53}_{-0.15-0.06-0.39}$
&$5.64^{+0.18+0.21+0.87}_{-0.18-0.10-0.59}$
&$0.54^{+0.45}_{-0.46}$~\cite{Cheng:2013fba} &$-24.2^{+0.2}_{-0.0}$~\cite{Zhang:2010zbud} & \\
$\bar B_d^0 \to  \bar K_0^{*}(1430)^0  \omega $ &$-3.30^{+0.10+0.09+0.44}_{-0.10-0.19-0.63}$
&$-4.24^{+0.13+0.23+0.73}_{-0.13-0.35-1.18}$
&$0.03^{+0.37}_{-0.35}$~\cite{Cheng:2013fba} &$10.0^{+0.1}_{-0.0}$~\cite{Zhang:2010zbud} &$-7\pm9$ \\
$\bar B_d^0 \to  \bar K_0^{*}(1430)^0  \phi $ &$0.15^{+0.00+0.06+0.20}_{-0.00-0.03-0.14}$
&$0.03^{+0.00+0.10+0.40}_{-0.00-0.09-0.30}$
&$0.43^{+0.04}_{-0.04}$~\cite{Cheng:2013fba} &&$12.4\pm8.1$\\
\hline
$\bar B_s^0 \to \bar K_0^{*}(1430)^0 K^{*0} $ &$0.68^{+0.02+0.00+0.04}_{-0.02-0.01-0.06}$
&$0.67^{+0.02+0.05+0.06}_{-0.02-0.05-0.07}$
&&& \\
$\bar B_s^0 \to  K_0^{*}(1430)^0 \bar K^{*0} $ &$0.20^{+0.01+0.05+0.17}_{-0.01-0.02-0.13}$
&$0.54^{+0.02+0.10+0.20}_{-0.02-0.09-0.20}$
&&& \\
$\bar B_s^0 \to  K_0^{*}(1430)^- K^{*+} $ &$0.41^{+0.01+0.09+0.17}_{-0.01-0.19-0.21}$
&$0.15^{+0.00+0.13+0.45}_{-0.00-0.28-0.54}$
&&$1.7^{+0.3}_{-0.4}$~\cite{Liu:2013lxz} & \\
$\bar B_s^0 \to K_0^{*}(1430)^+ K^{*-} $ &$-46.13^{+1.54+0.32+2.53}_{-1.49-0.31-1.11}$
&$-36.91^{+1.47+3.09+12.56}_{-1.44-2.53-10.14}$
&&$-63.6^{+10.7}_{-8.4}$~\cite{Liu:2013lxz} & \\
$\bar B_s^0 \to K_0^{*}(1430)^+ \rho^- $ &$15.11^{+0.63+0.87+4.48}_{-0.61-1.61-4.37}$
&$6.85^{+0.29+0.67+2.92}_{-0.28-0.94-2.86} $
&&$12.6^{+0.0}_{-0.0}$~\cite{Zhang:2010zbs} & \\
$\bar B_s^0 \to  K_0^{*}(1430)^0  \rho^0 $ &$-7.83^{+0.25+2.14+8.79}_{-0.25-1.02-6.17}$
&$8.98^{+0.29+3.37+15.96}_{-0.28-3.47-11.94}$
&&$84.5^{+0.1}_{-0.1}$~\cite{Zhang:2010zbs} & \\
$\bar B_s^0 \to  K_0^{*}(1430)^0  \omega $ &$-5.01^{+0.22+1.50+8.30}_{-0.24-3.11-11.42}$
&$-26.89^{+1.14+2.24+13.59}_{-1.24-3.71-17.23}$
&&$-86.7^{+0.1}_{-0.1}$~\cite{Zhang:2010zbs} & \\
$\bar B_s^0 \to  K_0^{*}(1430)^0  \phi $ &$-13.11^{+0.50+0.44+0.93}_{-0.49-0.20-0.80}$
&$-11.48^{+0.43+1.27+1.49}_{-0.42-1.14-1.30}$
&&& \\
\hline\hline
\end{tabular}
\end{center}
\label{tab:BSVCP}
\end{table}

Using the values of end-point parameters given above, we then present our numerical results and discussions.  The branching fractions and the direct $CP$ asymmetries of $ B_{u,d,s}\to K_0^{*}(1430)P\,(P=\pi,K) $ and $K_0^{*}(1430)V\,(V=K^*\,, \rho\,, \omega\,, \phi) $   decays  are given in tables~\ref{tab:BSP}, \ref{tab:BSV}, \ref{tab:BSPCP}, \ref{tab:BSVCP}, \ref{tab:bbpm} and~\ref{tab:bbcp}. For each result, the first error  is caused by the uncertainties of CKM parameters, the second error comes from the variation of quark masses, and the third error from the decay constants, form factors and Gegenbauer moments. In each table, the experimental data given by PDG~\cite{Zyla:2020p} or HFAG~\cite{Amhis:2019ckw} and the theoretical results obtained in previous works are also listed for comparison.
\begin{itemize}
\item
From tables~\ref{tab:BSP} and \ref{tab:BSV}, it can be found that  the branching fractions obtained based on S2 are generally much larger than the ones based on S1 because relatively larger decay constant of $K_0^{*}(1430)$ and  form factors of $B\to K_0^{*}(1430)$ transition are predicated in S2. In addition, most of predictions  based on S2~(S1) are favored~(disfavored)  by experimental data, the only exception is  $B^- \to  K_0^{*}(1430)^0 K^- $ decay. Our result  ${\cal B}(B^- \to  K_0^{*}(1430)^0 K^-) =(3.19^{+0.18+0.36+1.73}_{-0.14-0.62-1.39})\times 10^{-6}$ in S2 is consistent with the results obtained in the previous works, $(3.37^{+1.03}_{-0.85})\times 10^{-6}$~(QCDF)~\cite{Li:2015zra} and $(3.99\pm1.38)\times 10^{-6}$~(pQCD)~\cite{Wang:2020wcm}, however all of these  theoretical predictions are an order of magnitude larger  than experimental data $(0.38\pm0.13)\times 10^{-6}$.  The reason will be analyzed in the next item.
\item
$B^- \to K_0^{*}(1430)^0 K^-$ and   $K_0^{*}(1430)^- K^0$ decays are penguin dominated. After neglecting the power suppressed contribution, their simplified amplitudes can be written as
\begin{align}
&{\cal A}(B^- \to K_0^{*}(1430)^0 K^-)\sim (a_4^p-\gamma_\chi^{K_0^{*0}} a_6^p) \, A_{K^-  K_0^{*}(1430)^0 } \,,\\
&{\cal A}(B^- \to  K_0^{*}(1430)^- K^0)\sim (a_4^p-\gamma_\chi^{K^0} a_6^p)\, A_{K_0^{*}(1430)^0 K^-  } \,.
\end{align}
A significant feature of scalar meson is that its chiral factor is proportional to $M_S^2/(m_1(\mu)-m_2(\mu))$, which results in that $\gamma_\chi^{K_0^{*0}}$ is much larger than $\gamma_\chi^{K^0}$. Numerically, we obtain  $\gamma_\chi^{K_0^{*0}}:\gamma_\chi^{K^0}\approx12.6:1.4$ at $\mu\sim m_b$. As a result, it is expected that ${\cal B}(B^- \to K_0^{*}(1430)^0 K^-)\gg {\cal B}(B^- \to  K_0^{*}(1430)^- K^0)$. Moreover, for all of the penguin dominated $B\to SP$ and $SV$ decays, the decay modes with $M_2=S$~($M_2$ is the emitted meson) generally have relatively larger branching fractions than the ones with  $M_2=(P,V)$, which can also be found from the numerical results listed in  tables~\ref{tab:BSP} and \ref{tab:BSV}.

In addition, the large $\gamma_\chi^{K_0^{*0}}$ also results in the large  theoretical predictions for ${\cal B}(B^- \to  K_0^{*}(1430)^0 K^-)$ compared with data, which has been mentioned in the last item. It should be noted that the significance of data,  $(0.38\pm0.13)\times 10^{-6}$, is smaller than $3\sigma$, thus more precise measurement on ${\cal B}(B^- \to  K_0^{*}(1430)^0 K^-)$ is required for confirming or refuting such possible anomaly.

\item
The $\bar B_d^0 \to  K_0^{*}(1430)^- K^{(*)+} $ and $K_0^{*}(1430)^+ K^{(*)-}$ decays are pure annihilation processes, and thus are very suitable for probing the effects of annihilation corrections. However, these decays have very small  branching fractions $\sim 10^{-7}$ because their amplitudes are power suppressed, and therefore are not easy to be precisely measured in the near future.

\item
The $\bar B_s^0 \to K_0^{*}(1430)^+ \pi^-/\rho^- $ and $\bar B_s^0 \to K_0^{*}(1430)^0 \pi^0/ \rho^0(\omega)$ decays are tree-dominated, while the former are color-allowed and the later are color-suppressed. As a result, the branching fractions of former are an order of magnitude larger than the later.

\item
The $SU(3)$ flavor symmetry indicates some useful relations between the decays considered in this work. Taking $B_{u,d,s}\to SP$ decays as examples, after applying   $SU(3)$ flavor symmetry on the spectator quark, one may expect that
\begin{align}
{\cal A}(B^- \to  K_0^{*}(1430)^- K^0 ) \approx & {\cal A}(\bar B_d^0 \to \bar K_0^{*}(1430)^0 K^0 )  \, ,\\
{\cal A}(B^- \to K_0^{*}(1430)^0 K^- ) \approx & {\cal A}(\bar  B_d^0 \to  K_0^{*}(1430)^0 \bar K^0) \, ,\\
{\cal A}(B^- \to \bar K_0^{*}(1430)^0 \pi^- ) \approx & {\cal A}(\bar B_d^0 \to  \bar K_0^{*}(1430)^0  \pi^0) \approx  {\cal A}(\bar B_s^0 \to \bar K_0^{*}(1430)^0 K^0) \, ,\\
{\cal A}(B^- \to  K_0^{*}(1430)^- \pi^0 ) \approx & {\cal A}(B_d^0 \to K_0^{*}(1430)^- \pi^+ ) \approx  {\cal A}(\bar B_s^0 \to  K_0^{*}(1430)^- K^+) \, ,
\end{align}
which further imply that
\begin{align}
R_1 &\equiv \frac{ \Gamma_{B^- \to  K_0^{*}(1430)^- K^0 } }{ \Gamma_{\bar B_d^0 \to \bar K_0^{*}(1430)^0 K^0 }}\approx 1  \, ,\qquad
R_2  \equiv \frac{  \Gamma_{B^- \to K_0^{*}(1430)^0 K^- } } {\Gamma_{\bar  B_d^0 \to  K_0^{*}(1430)^0 \bar K^0) }}\approx 1 \, ,\\
R_3 &\equiv \frac{\Gamma_{B^- \to \bar K_0^{*}(1430)^0 \pi^- } }{ 2\Gamma_{\bar B_d^0 \to  \bar K_0^{*}(1430)^0  \pi^0}} \approx 1 \, ,\qquad
R_3' \equiv\frac{\Gamma_{B^- \to \bar K_0^{*}(1430)^0 \pi^- } }{ \Gamma_{\bar B_s^0 \to \bar K_0^{*}(1430)^0 K^0} }\approx 1 \, ,\\
R_4 &\equiv \frac{2 \Gamma_{B^- \to  K_0^{*}(1430)^- \pi^0 } }{ \Gamma_{B_d^0 \to K_0^{*}(1430)^- \pi^+ } } \approx 1 \, ,\qquad
R_4' \equiv \frac{2 \Gamma_{B^- \to  K_0^{*}(1430)^- \pi^0 }}{\Gamma_{\bar B_s^0 \to  K_0^{*}(1430)^- K^+} }\approx 1 \, .
\end{align}
Besides, the  $SU(3)$ flavor symmetry also expects  that
\begin{align}
R_5  \equiv \frac{  \Gamma_{\bar B_d^0 \to K_0^{*}(1430)^- \pi^+ } } {2 \Gamma_{\bar B_d^0 \to  \bar K_0^{*}(1430)^0  \pi^0}}\approx 1 \, ,\qquad
R_6  \equiv \frac{\Gamma_{B^- \to \bar K_0^{*}(1430)^0 \pi^- } }{ 2\Gamma_{ B^- \to  K_0^{*}(1430)^- \pi^0}} \approx 1 \, .
\end{align}
It is found in our calculation~(S2) that
\begin{align}
R_1 &= 0.93  \, ,\quad
R_2  =1.00 \, ,\quad
R_3 =1.12 \, ,\quad
R_3' = 1.10 \, ,\quad
R_4 =0.99 \, ,\quad
R_4' =1.08 \,,\\
R_5&= 1.11  \, ,\quad
R_6  =0.90\,,
\end{align}
which generally agree with the expectations of  $SU(3)$ flavor symmetry. The flavor symmetry breaking effect is mainly ascribed to the remaining suppressed contributions.  Taking $R_6$ as an example, the amplitude of $B^- \to  K_0^{*}(1430)^- \pi^0$ decay receives additional CKM-suppressed contributions proportional to $\delta_{pu}\alpha_1 $, as well as  CKM- and color-suppressed  $\delta_{pu}\alpha_2$, compared with with $B^- \to \bar K_0^{*}(1430)^0 \pi^-$ decay.

\item { The ${\cal O}(\alpha_s^2)$ corrections to the amplitudes of non-leptonic two-body $B$ decays have been evaluated in recent years~\cite{Beneke:2009ek,Bell:2015koa,Bell:2020qus,Huber:2016xod,Bell:2007tv,Bell:2009nk,Beneke:2005vv,Kivel:2006xc,Pilipp:2007mg,Beneke:2006mk,Kim:2011jm}. In Ref.~\cite{Beneke:2009ek}, it is found that the  next-to-next-to-leading order (NNLO) vertex correction to the color-suppressed amplitude $\alpha_2$ is  sizable, but when  combined with the ${\cal O}(\alpha_s^2)$  correction to spectator scattering, the overall NNLO corrections to the color-allowed and -suppressed tree amplitudes are small due to the large cancellation. Therefore, the ${\cal O}(\alpha_s^2)$ corrections to the tree-dominated $\bar B_s^0 \to K_0^{*}(1430)^+ \pi^-/\rho^- $ and $\bar B_s^0 \to K_0^{*}(1430)^0 \pi^0/ \rho^0(\omega)$ decays would not be significant. The NNLO correction to the penguin amplitude has also been studied in Refs.~\cite{Bell:2015koa,Bell:2020qus}. It is found that the NNLO contributions from current-current and penguin operators are sizable, but there is a strong cancellation between them, which results in a much reduced overall NNLO corrections to the penguin amplitude $a_4^{u,c}$. As a consequence the full NNLO result for $a_4^{u,c}$ is very close to the NLO result~\cite{Bell:2020qus}~(an example, $a_4^{u,c}(\pi \bar{K})$, is shown by Fig.~8 in Ref.~\cite{Bell:2020qus} ). Therefore, based on these previous works on the  ${\cal O}(\alpha_s^2)$ correction, we can expect that the ${\cal O}(\alpha_s^2)$ corrections do not affect the main findings of this work. }

\end{itemize}

  \begin{table}[t]
  \caption{The branching fractions (in units of ${10}^{-6}$) of  ${\bar B}^0_{d,s}\to \bar{K}_0^*(1430)  {K}^{(*)}+ c.c.$  decays  in S2.}
  \begin{center}
  \begin{tabular}{lccc}
  \hline\hline
   Decay modes   & this work &pQCD\cite{Liu:2013lxz} &data  \\\hline
 $\bar B_d^0 \to K_0^{*0}\bar{K}^{0}\,\,+c.c.$ &$3.05^{+0.17+0.35+1.61}_{-0.13-0.59-1.29} $\\
 ${\bar B}^0_{d}\to K_0^{*+}K^{-}+c.c.$ &$0.17^{+0.01+0.03+0.07}_{-0.01-0.05-0.05}$\\
 ${\bar B}^0_{s}\to K_0^{*0}\bar{K}^{0}\,\,+c.c.$ &$32.78^{+1.45+4.17+16.59}_{-1.02-7.08-13.39}$
 &&$33.0\pm10.1$\\
 ${\bar B}^0_{s}\to K_0^{*+}K^{-}+c.c.$ &$33.73^{+1.51+4.17+16.65}_{-1.07-7.09-13.43}$
 &&$31.3\pm25.4$\\
  \hline
 $\bar B_d^0 \to K_0^{*0}\bar{K}^{*0}\,\,+c.c.$ &$2.52^{+0.14+0.29+1.28}_{-0.11-0.49-1.03}$ &$0.59^{+0.1}_{-0.1}$ & \\
 ${\bar B}^0_{d}\to K_0^{*+}K^{*-}+c.c.$ &$0.07^{+0.01+0.01+0.03}_{-0.00-0.01-0.02}$
 &$1.1^{+0.1}_{-0.1}$ & \\
 ${\bar B}^0_{s}\to K_0^{*0}\bar{K}^{*0}\,\,+c.c.$ &$34.98^{+1.55+3.56+18.02}_{-1.09-6.14-14.57}$
 &$13.0^{+2.0}_{-2.0}$ & \\
 ${\bar B}^0_{s}\to K_0^{*+}K^{*-}+c.c.$ &$33.81^{+1.51+3.47+17.13}_{-1.07-5.98-13.85}$
 &$15.0^{+4.0}_{-3.0}$ & \\
  \hline\hline
  \end{tabular}
  \label{tab:bbpm}
  \end{center}
  \end{table}

\begin{table}[t]
  \caption{The $CP$ asymmetry parameters (in units of ${10}^{-2}$) of
  ${\bar B}^0_{d,s}\to \bar{K}_0^*(1430)  {K}^{(*)}+ c.c.$ decays in S2.}
  \begin{tabular}{lccccc}
  \hline\hline
     &$A_{CP}$ &C &${\Delta} C$ &S  & ${\Delta} S $ \\\hline
 ${\bar B}^0_{d}\to \bar K_0^{*0} {K}^{0}+c.c.$ &$-20.24^{+0.77}_{-0.75}$ &$10.04^{+0.37}_{-0.38}$
 &$-10.72^{+0.41}_{-0.40}$ &$2.15^{+0.08}_{-0.08}$ &$8.70^{+0.36}_{-0.36}$ \\
 ${\bar B}^0_{d}\to K_0^{*-}K^{+}+c.c.$ &$1.42^{+0.06}_{-0.05}$ &$-0.04^{+0.00}_{-0.00}$
 &$1.42^{+0.05}_{-0.05}$ &$4.76^{+8.13}_{-7.64}$ &$6.35^{+0.22}_{-0.25}$ \\
 $\bar B_s^0 \to \bar K_0^{*0} {K}^{0}+c.c.$ &$-0.74^{+0.02}_{-0.02}$ &$-0.42^{+0.01}_{-0.01}$
 &$-0.40^{+0.01}_{-0.01}$ &$0.03^{+0.00}_{-0.00}$ &$0.31^{+0.01}_{-0.01}$ \\
 ${\bar B}^0_{s}\to K_0^{*-}K^{+}+c.c.$ &$-8.56^{+0.26}_{-0.26}$ &$31.27^{+0.76}_{-0.78}$
 &$-32.52^{+0.82}_{-0.80}$ &$-11.39^{+0.60}_{-0.56}$ &$3.49^{+0.50}_{-0.54}$\\
 \hline
 ${\bar B}^0_{d} \to \bar K_0^{*0} {K}^{*0}+c.c.$ &$-14.03^{+0.53}_{-0.52}$ &$9.87^{+0.36}_{-0.37}$
 &$-5.91^{+0.23}_{-0.22}$ &$-14.20^{+0.54}_{-0.53}$ &$-6.58^{+0.24}_{-0.23}$\\
 ${\bar B}^0_{d}\to K_0^{*-}K^{*+}+c.c.$ &$-0.52^{+0.02}_{-0.02}$ &$4.69^{+0.18}_{-0.18}$
 &$0.42^{+0.02}_{-0.02}$ &$5.61^{+8.16}_{-7.67}$ &$-0.75^{+0.03}_{-0.03}$\\
 $\bar B_s^0 \to \bar K_0^{*0} {K}^{*0}+c.c.$ &$-0.55^{+0.02}_{-0.02}$ &$-0.44^{+0.01}_{-0.01}$
 &$-0.24^{+0.01}_{-0.01}$ &$0.70^{+0.02}_{-0.02}$ &$-0.23^{+0.01}_{-0.01}$\\
 ${\bar B}^0_{s}\to K_0^{*-}K^{*+}+c.c.$ &$-7.74^{+0.23}_{-0.23}$ &$22.86^{+0.74}_{-0.77}$
 &$-23.27^{+0.77}_{-0.75}$ &$30.72^{+0.69}_{-0.71}$ &$-23.55^{+0.49}_{-0.47}$\\
  \hline\hline
  \end{tabular}
  \label{tab:bbcp}
  \end{table}

Different from the other decay modes, ${\bar B}^0_{s}$ or ${\bar B}^0_{d}$  can decay into  $\bar{K}_0^*(1430)  {K}^{(*)}$ and its CP conjugate state simultaneously. The sum of the $\bar{K}_0^*(1430)  {K}^{(*)}$ and $K_0^*(1430)  \bar{K}^{(*)}$ decay rates is measured more accurately than the individual rates. Our prediction for ${\bar B}^0_{d,s}\to \bar{K}_0^*(1430)  {K}^{(*)}+ c.c.$ decays are given in table~\ref{tab:bbpm}, {the LHCb data~\cite{Aaij:2019nmr} given by Eq.~\eqref{Bskk} and the pQCD predictions~\cite{Liu:2013lxz} are also listed for comparision.} In addition, the CP violations of  ${\bar B}^0_{d,s}\to \bar{K}_0^*(1430)  {K}^{(*)}+ c.c.$ decays are much more complicated because $\bar{K}_0^*(1430)  {K}^{(*)}+ c.c.$ are not CP eigenstates. The system of these decays define five CP asymmetry parameters, $C$, ${\Delta} C$, $S$, ${\Delta} S $ and $A_{CP}$, where $S$ is referred to as mixing-induced CP asymmetry,  $C$ is the direct CP asymmetry, ${\Delta} C$ and ${\Delta} S$ are CP conserving
quantities, and $A_{CP}$ is the time-integrated charge asymmetry. Our results for these observables are first given in table~\ref{tab:bbcp}.

The  ${\bar B}^0_{d}\to K_0^{*-}K^{(*)+}+c.c.$ decays are caused by the pure annihilation transition, and therefore have very small branching fractions $\sim {\cal O}(10^{-7})$.  The  ${\bar B}^0_{d} \to \bar K_0^{*0} {K}^{(*)0}+c.c.$  decays are dominated by penguin contributions, but are CKM-suppressed.   The penguin-dominated  ${\bar B}^0_{s}\to K_0^{*-}K^{(*)+}+c.c.$ and $\bar K_0^{*0} {K}^{(*)0}+c.c.$  decays have relatively large branching fractions, and our results for the $SP$ modes are in good agreement with the data obtained by the LHCb collaboration with significance over $10$ standard deviations~\cite{Aaij:2019nmr}. However,  some systematic variations do impact
strongly on the need to include tensor resonances in the fit model, and thus the model uncertainties of experimental data are very large~\cite{Aaij:2019nmr}. It is expected that the future refined measurements for the ${\bar B}^0_{d,s}\to \bar{K}_0^*(1430)  {K}^{(*)}+ c.c.$ decays can provide serious test on the theoretical predictions.


\section{Summary}
In this paper, the nonleptonic charmless $B_{u,d,s}  \to K_0^*(1430)P$~$(P=K\,, \pi)$ and $ K_0^*(1430)V$~$ (V=K^*\,, \rho\,, \omega\,, \phi)$ decays are studied. The amplitudes are calculated by using the QCD factorization approach, and the non-perturbative quantities~(form factor, decay constant and distribution amplitudes) are evaluated by using a covariant light-front approach.  The branching fractions and CP asymmetries of theses decay modes are calculated, our main theoretical results are collected in tables \ref{tab:BSP}-\ref{tab:bbcp}. Some decay modes are first predicted in this work. In order to test the underlying structure of $K_0^*(1430)$ meson, our calculation are made based on two different scenarios that   $K_0^*(1430)$ is the first excited~(scenario S1)  and the lowest-lying~(scenario S2) p-wave two-quark state. Comparing our results with data, we find that the scenarios S1 and S2  lead to significantly different results for branching factions, and the results based on scenario S2  agree well with the experimental data, which implies that $K_0^*(1430)$ as the lowest-lying~(scenario S2) p-wave $(s,u/d)$ state is favored by data. However, the theoretical results for ${\cal B}(B^- \to  K_0^{*}(1430)^0 K^-)$ are much larger than data due to the large chiral factor $\gamma_\chi^{K_0^{*0}}$. The future refined measurement on ${\cal B}(B^- \to  K_0^{*}(1430)^0 K^-)$ is required for confirming or refuting such possible anomaly. Besides, some useful relations based on $SU(3)$ flavor symmetry are studied.

\section*{Acknowledgements}
This work is supported by the National Natural Science Foundation of China (Grant No. 11875122), Excellent Youth Foundation of Henan Province (Grant No. 212300410010),  The Youth Talent Support Program of Henan Province (Grant No. ZYQR201912178) and the Program for Innovative Research Team in University of Henan Province (Grant No. 19IRTSTHN018).

\section*{Appendix A: Decay constant, form factor and DAs }
The decay constant, form factor and DAs are essential nonperturbative inputs for the amplitudes of $B\to SP$ and $SV$. In this sections, we would like to clarify our convention for the definitions of these nonperturbative quantities.
\subsection*{A.1 Decay constant}
The decay constants of pseudoscalar~(P), vector~(V) and scalar~(S) mesons are defined as
\begin{align}\label{f}
\langle P(p)|\bar{q}_1\gamma^{\mu}\gamma^{5} q_2|0\rangle &=-if_{P}p_\mu\,,\quad
\langle V(p,\epsilon)|\bar{q}_1\gamma^{\mu} q_2|0\rangle =-if_{V}m_V\epsilon^{*\mu}\,,\nonumber \\
\langle S(p)|\bar{q}_1\gamma^{\mu} q_2|0\rangle &=f_{S}p_\mu\,,\qquad
\langle S(p)|\bar{q}_1 q_2|0\rangle = m_S\bar f_{S}(\mu)\,.
\end{align}
The scale-dependent scalar decay constant $\bar f_S(\mu)$ and the vector
decay constant $f_{S}$ are related by the equation of motion, and thus have the following relation,
\begin{align}\label{eq:fsfsbar}
\bar f_S=\frac{\mu_S}{m_S}f_S\equiv \bar{\mu}_S f_S \,,\qquad \text{with}\qquad \mu_S=\frac{m_S^2}{m_1(\mu)-m_2(\mu)}\,,
\end{align}
where $m_1(\mu)$ and $m_2(\mu)$ are running masses of current quarks.


\subsection*{A.2 Form factor}
The  form factors  of $B\to S\,,P\,,V$ transitions concerned in this paper can be defined as
\begin{align}\label{FF}
{\langle}S(p^{\prime\prime})|{\bar q}\gamma_{\mu}\gamma_{5}b|B(p^\prime){\rangle}
=&-i\bigg[\bigg(P_\mu-\frac{m_B^2-m_S^2}{q^2}q_\mu\bigg
)U_1(q^2)+\frac{m_B^2-m_S^2}{q^2}q_\mu U_0(q^2)\bigg]\,,
\\
{\langle}P(p^{\prime\prime})|{\bar q}\gamma_{\mu}b|B(p^\prime){\rangle}
=&\bigg(P_\mu-\frac{m_B^2-m_P^2}{q^2}q_\mu\bigg)F_1(q^2)
+\frac{m_B^2-m_P^2}{q^2}q_\mu F_0(q^2)\,,
\\
{\langle}V(\epsilon,p^{\prime\prime})|{\bar q}\gamma_{\mu}b|B(p^\prime){\rangle}
=&-i\frac{V(q^2)}{m_B+m_V}\varepsilon_{\mu\nu\alpha\beta} \epsilon^{*\nu}P^{\alpha}q^{\beta}\,,
\\
{\langle}V(\epsilon,p^{\prime\prime})|{\bar q}\gamma_{\mu}\gamma_{5}b|B(p^\prime){\rangle}
=&2m_V\frac{\epsilon^*\cdot P}{q^2}q_\mu A_0(q^2)+(m_B+m_V)\bigg(\epsilon_\mu^*-\frac{\epsilon^*\cdot P}{q^2}q_\mu\bigg)A_1(q^2)\nonumber \\
&-\frac{\epsilon^*\cdot P}{m_B+m_V}\bigg(P_\mu-\frac{m_B^2-m_V^2}{q^2}q_\mu\bigg)A_2(q^2)\,,
\end{align}
where $P_\mu=(p^\prime+p^{\prime\prime}),\,q_\mu=(p^\prime-p^{\prime\prime})$ and $\varepsilon_{0123}=-1$.
The momentum dependence of form factor can be parameterized as
\begin{eqnarray}\label{eq:dipo}
F(q^2)=\frac{F(0)}{1-a\,(q^2/m^2_{ B})+b\,(q^2/m^2_{ B})^2}\,.
\end{eqnarray}
The values of parameters $a$, $b$ and $F(0)$ for the transition concerned in this work are summarized in Table~\ref{tab:BSPSVFF}.


 \subsection*{A.2 Distribution amplitude}
 For the light-cone distribution amplitudes of pseudoscalar and vector mesons, we take the same definition and convention used in the Ref.~\cite{Beneke:2003zv}.
For the scalar meson, the twist-2 and -3 light-cone distribution amplitudes are given by
\begin{align}\label{DAfs}
\langle S(p)|{\bar{q}_2}(z_2)\gamma_{\mu} q_1(z_1)|0\rangle &=f_{S}p_\mu\int_0^1dx e^{i(xp\cdot z_2+\bar x p\cdot z_1)}\Phi_S(x),
\\
\langle S(p)|{\bar{q}_2}(z_2) q_1(z_1)|0\rangle &=f_{S}\mu_S\int_0^1dx e^{i(xp\cdot z_2+\bar x p\cdot z_1)}\phi_S(x)\,,\\
\langle S(p)|{\bar{q}_2}(z_2)\sigma_{\mu\nu} q_1(z_1)|0\rangle &=-f_{S}\mu_S(p_\mu z_\nu-p_\nu z_\mu)\int_0^1dx e^{i(xp\cdot z_2+\bar x p\cdot z_1)}\frac{\phi_S^\sigma(x)}{6}\,.
\end{align}



The leading-twist DAs are conventionally expended in Gegenbauer polynomials,
\begin{align}\label{eq:tw2Phi}
\Phi_M(x,\mu)=&6x(1-x)\left[1+\sum^{\infty}_{n=1} a_n^M(\mu)C_n^{3/2}(x-\bar x) \right],
\end{align}
where $M=P\,, S\,, V$. Equivalently, for the scalar meson, one can also use  
\begin{align}
\label{eq:tw2PhiS}
\Phi_S(x)=&6x(1-x)\bar \mu_S(\mu)\left[b_0^S+\sum^{\infty}_{n=1}b_n^S(\mu)C_n^{3/2}(x-\bar x) \right],
\end{align}
where, $ a_n^S(\mu)={\bar \mu_S}(\mu) b_n^S (\mu)$ and $\bar \mu_S(\mu)$ can be conveniently absorbed by  $f_S$ via Eq.~\eqref{eq:fsfsbar}.


When three-particle contributions are neglected, the twist-3 two-particle DAs of P and S mesons are determined completely by the equation of motion, which then require
\begin{align}
\label{eq:tw3Phis}
\phi_{P,S}=1\,,\qquad \phi_{P,S}^{\sigma}=6x\bar{x}\,;
\end{align}
For the vector meson,  the twist-3 DAs can be expressed in terms of the leading-twist DA $\Phi_\bot(x)$ of transversely polarized state,  are usually expressed by the function, $\phi_V$, defined as
\begin{align}
\label{eq:tw3Phiv}
\phi_V(x,\mu)\equiv &\int_0^x dv \frac{\Phi_\perp(v)}{\bar v}-\int_x^1 dv \frac{\Phi_\perp(v)}{ v}
=3\sum_{n=0}^\infty a_{n,\perp}^{V}(\mu)P_{n+1}(2x-1),
\end{align}
where $a_{0,\perp}^{V}=1$ and $P_n(x)$ are the Legendre polynomials.
\section*{Appendix B: Decay amplitudes}
The amplitudes of $B_{u,d,s}\to K_0^*(1430)P$~$(P=K\,, \pi)$ decays are written as
\begin{align}
{\cal A}_{ B^- \to  K_0^{*-} K^0}=&A_{K_0^{*} K}[\delta_{pu}\beta_2+\alpha_4^p-\frac12\alpha_{4,EW}^p+\beta_3^p+\beta_{3,EW}^p]\,,\\
{\cal A}_{ B^- \to  K_0^{*0} K^-}=&A_{K K_0^{*} }[\delta_{pu}\beta_2+\alpha_4^p-\frac12\alpha_{4,EW}^p+\beta_3^p+\beta_{3,EW}^p]\,,\\
{\cal A}_{ B^- \to \bar K_0^{*0} \pi^-}=&A_{\pi K_0^*}[\delta_{pu}\beta_2+\alpha_4^p-\frac12\alpha_{4,EW}^p+\beta_3^p+\beta_{3,EW}^p]\,,\\
{\cal A}_{ B^- \to  K_0^{*-} \pi^0}=&\frac{1}{\sqrt2}A_{\pi K_0^{*}}[\delta_{pu}(\alpha_1+\beta_2)+\alpha_4^p+\alpha_{4,EW}^p+\beta_3^p+\beta_{3,EW}^p]\nonumber\\
&+\frac{1}{\sqrt2}A_{K_0^{*}\pi}[\delta_{pu}\alpha_2+\frac32\alpha_{3,EW}^p]\,,
\end{align}
\begin{align}
{\cal A}_{\bar B^0 \to \bar K_0^{*0} K^0}=&A_{ K_0^{*}K}[\alpha_4^p-\frac{1}{2}\alpha_{4,EW}^p+\beta_3^p+\beta_4^p-\frac{1}{2}\beta_{3,EW}^p-\frac{1}{2}\beta_{4,EW}^p]\nonumber\\
&+B_{K K_0^{*} }[b_4^p-\frac{1}{2}b_{4,EW}^p] \,,  \\
{\cal A}_{\bar B^0 \to  K_0^{*0} \bar K^0}=&A_{K K_0^*}[\alpha_4^p-\frac{1}{2}\alpha_{4,EW}^p+\beta_3^p+\beta_4^p-\frac{1}{2}\beta_{3,EW}^p-\frac{1}{2}\beta_{4,EW}^p]\nonumber\\
&+B_{K_0^*  K }[b_4^p-\frac{1}{2}b_{4,EW}^p] \,,  \\
{\cal A}_{\bar B^0 \to  K_0^{*-}K^+}=&A_{K_0^* K}[\delta_{pu}\beta_1+\beta_4^p+\beta_{4,EW}^p]+B_{K K_0^*}[b_4^p
-\frac{1}{2}b_{4,EW}^p] \,,   \\
{\cal A}_{\bar B^0 \to  K_0^{*+} K^-}=&A_{ K K_0^*}[\delta_{pu}\beta_1+\beta_4^p+\beta_{4,EW}^p]+B_{K_0^* K}[b_4^p-\frac{1}{2}b_{4,EW}^p] \,,   \\
{\cal A}_{ \bar B^0 \to  K_0^{*-} \pi^+}=&A_{\pi K_0^*}[\delta_{pu}\alpha_1+\alpha_4^p+\alpha_{4,EW}^p+\beta_3^p-\frac12\beta_{3,EW}^p]\,,\\
 {\cal A}_{\bar B^0 \to \bar K_0^{*0} \pi^0}=&\frac{1}{\sqrt2}A_{\pi K_0^*}[-\alpha_4^p+\frac12\alpha_{4,EW}^p-\beta_3^p+\frac12\beta_{3,EW}^p]\nonumber\\
&+\frac{1}{\sqrt2}A_{K_0^*\pi}[\delta_{pu}\alpha_2+\frac32\alpha_{3,EW}^p]\,,
\end{align}
\begin{align}
{\cal A}_{\bar B_s^0 \to \bar K_0^{*0}K^0}=&A_{K K_0^*}[\alpha_4^p-\frac{1}{2}\alpha_{4,EW}^p+\beta_3^p+\beta_4^p-\frac{1}{2}\beta_{3,EW}^p-\frac{1}{2}\beta_{4,EW}^p] \nonumber\\
&+B_{ K_0^* K}[b_4^p-\frac{1}{2}b_{4,EW}^p]\,,  \\
{\cal A}_{\bar B_s^0 \to  K_0^{*0} \bar K^0}=&A_{K_0^* K}[\alpha_4^p-\frac{1}{2}\alpha_{4,EW}^p+\beta_3^p+\beta_4^p-\frac{1}{2}\beta_{3,EW}^p-\frac{1}{2}\beta_{4,EW}^p] \nonumber\\
&+B_{ K K_0^*}[b_4^p-\frac{1}{2}b_{4,EW}^p]\,,  \\
{\cal A}_{\bar B_s^0 \to  K_0^{*-}K^+}=&A_{K K_0^*}[\delta_{pu}\alpha_1+\alpha_4^p+\alpha_{4,EW}^p+\beta_3^p+\beta_4^p-\frac{1}{2}\beta_{3,EW}^p-\frac{1}{2}\beta_{4,EW}^p] \nonumber \\
&+B_{ K_0^* K}[\delta_{pu}b_1+b_4^p+b_{4,EW}^p]\,,  \\
{\cal A}_{\bar B_s^0 \to  K_0^{*+}K^-}=&A_{K_0^* K}[\delta_{pu}\alpha_1+\alpha_4^p+\alpha_{4,EW}^p+\beta_3^p+\beta_4^p
-\frac{1}{2}\beta_{3,EW}^p-\frac{1}{2}\beta_{4,EW}^p]\nonumber \\
& +B_{ K K_0^*}[\delta_{pu}b_1+b_4^p+b_{4,EW}^p]\,,   \\
{\cal A}_{\bar B_s^0 \to  K_0^{*+} \pi^-}=&A_{K_0^*\pi}[\delta_{pu}\alpha_1+\alpha_4^p+\alpha_{4,EW}^p+\beta_3^p
-\frac{1}{2}\beta_{3,EW}^p] \,,   \\
{\cal A}_{\bar B_s^0 \to  K_0^{*0} \pi^0}=&\frac{1}{\sqrt 2}A_{K_0^*\pi}[\delta_{pu}\alpha_2-\alpha_4^p+\frac32\alpha_{3,EW}^p+\frac12\alpha_{4,EW}^p-\beta_{3}^p
+\frac{1}{2}\beta_{3,EW}^p] \,.
\end{align}

The amplitudes of $B_{u,d,s}\to K_0^*(1430)V$~$(V=\rho\,,K^*\,, \omega\,,\phi)$ decays are written as
\begin{align}
{\cal A}_{ B^- \to  K_0^{*-} K^{*0}}=&A_{K_0^{*} K^{*}}[\delta_{pu}\beta_2+\alpha_4^p-\frac12\alpha_{4,EW}^p+\beta_3^p+\beta_{3,EW}^p]\,,\\
{\cal A}_{ B^- \to  K_0^{*0} K^{*-}}=&A_{K^{*} K_0^{*} }[\delta_{pu}\beta_2+\alpha_4^p-\frac12\alpha_{4,EW}^p+\beta_3^p+\beta_{3,EW}^p]\,,\\
{\cal A}_{ B^- \to \bar K_0^{*0} \rho^-}=&A_{\rho K_0^*}[\delta_{pu}\beta_2+\alpha_4^p-\frac12\alpha_{4,EW}^p+\beta_3^p+\beta_{3,EW}^p]\,,\\
 {\cal A}_{ B^- \to  K_0^{*-} \rho^0}=&\frac{1}{\sqrt2}A_{\rho K_0^*}[\delta_{pu}(\alpha_1+\beta_2)+\alpha_4^p+\alpha_{4,EW}^p+\beta_3^p+\beta_{3,EW}^p]\nonumber \\
&+\frac{1}{\sqrt2}A_{K_0^*\rho}[\delta_{pu}\alpha_2+\frac32\alpha_{3,EW}^p]\,,\\
{\cal A}_{ B^- \to  K_0^{*-} \omega}=&\frac{1}{\sqrt2}A_{\omega K_0^*}[\delta_{pu}(\alpha_1+\beta_2)+\alpha_4^p+\alpha_{4,EW}^p+\beta_3^p+\beta_{3,EW}^p]\nonumber \\
&+\frac{1}{\sqrt2}A_{K_0^*\omega}[\delta_{pu}\alpha_2+2\alpha_3^p+\frac12\alpha_{3,EW}^p] \,,\\
{\cal A}_{ B^- \to  K_0^{*-} \phi}=& A_{K_0^*\phi}[\delta_{pu}\beta_2+\alpha_3^p+\alpha_4^p-\frac12\alpha_{3,EW}^p
-\frac12\alpha_{4,EW}^p+\beta_3^p+\beta_{3,EW}^p]\,,
\end{align}
\begin{align}
{\cal A}_{\bar B^0 \to \bar K_0^{*0} K^{*0}}=&A_{ K_0^*K^{*}}[\alpha_4^p-\frac{1}{2}\alpha_{4,EW}^p+\beta_3^p+\beta_4^p-\frac{1}{2}\beta_{3,EW}^p-\frac{1}{2}\beta_{4,EW}^p]\nonumber \\
&+B_{K^{*}  K_0^* }[b_4^p-\frac{1}{2}b_{4,EW}^p] \,,  \\
{\cal A}_{\bar B^0 \to  K_0^{*0} \bar K^{*0}}=&A_{ K^{*} K_0^*}[\alpha_4^p-\frac{1}{2}\alpha_{4,EW}^p+\beta_3^p+\beta_4^p-\frac{1}{2}\beta_{3,EW}^p-\frac{1}{2}\beta_{4,EW}^p]\nonumber \\
&+B_{K_0^* K^{*} }[b_4^p-\frac{1}{2}b_{4,EW}^p] \,,  \\
{\cal A}_{\bar B^0 \to  K_0^{*-} K^{*+}}=&A_{ K_0^* K^{*}}[\delta_{pu}\beta_1+\beta_4^p+\beta_{4,EW}^p]+B_{K^{*} K_0^*}[b_4^p
-\frac{1}{2}b_{4,EW}^p] \,,   \\
{\cal A}_{\bar B^0 \to  K_0^{*+} K^{*-}}=&A_{K^{*} K_0^*}[\delta_{pu}\beta_1+\beta_4^p+\beta_{4,EW}^p]+B_{K_0^* K^{*}}[b_4^p-\frac{1}{2}b_{4,EW}^p] \,,   \\
{\cal A}_{ \bar B^0 \to  K_0^{*-} \rho^+}=&A_{\rho K_0^*}[\delta_{pu}\alpha_1+\alpha_4^p+\alpha_{4,EW}^p+\beta_3^p-\frac12\beta_{3,EW}^p]\,,\\
 {\cal A}_{\bar B^0 \to \bar K_0^{*0} \rho^0}=&\frac{1}{\sqrt2}A_{\rho K_0^*}[-\alpha_4^p+\frac12\alpha_{4,EW}^p-\beta_3^p+\frac12\beta_{3,EW}^p]\nonumber \\
&
+\frac{1}{\sqrt2}A_{K_0^*\rho}[\delta_{pu}\alpha_2+\frac32\alpha_{3,EW}^p]\,,\\
 {\cal A}_{ \bar B^0 \to  \bar K_0^{*0} \omega}=&\frac{1}{\sqrt2}A_{\omega K_0^*}[\alpha_4^p-\frac12\alpha_{4,EW}^p+\beta_3^p-\frac12\beta_{3,EW}^p]\nonumber \\
&
+\frac{1}{\sqrt2}A_{K_0^*\omega}[\delta_{pu}\alpha_2+2\alpha_3^p+\frac12\alpha_{3,EW}^p]\,, \\
{\cal A}_{ \bar B^0 \to  \bar K_0^{*0} \phi}=& A_{K_0^*\phi}[\alpha_3^p+\alpha_4^p-\frac12\alpha_{3,EW}^p
-\frac12\alpha_{4,EW}^p+\beta_3^p-\frac12\beta_{3,EW}^p]\,,
\end{align}
\begin{align}
{\cal A}_{\bar B_s^0 \to \bar K_0^{*0}K^{*0}}=&A_{K^* K_0^*}[\alpha_4^p-\frac{1}{2}\alpha_{4,EW}^p+\beta_3^p+\beta_4^p-\frac{1}{2}\beta_{3,EW}^p-\frac{1}{2}\beta_{4,EW}^p] \nonumber \\
&+B_{ K_0^* K^*}[b_4^p-\frac{1}{2}b_{4,EW}^p]\,,  \\
{\cal A}_{\bar B_s^0 \to  K_0^{*0} \bar K^{*0}}=&A_{K_0^* K^*}[\alpha_4^p-\frac{1}{2}\alpha_{4,EW}^p+\beta_3^p+\beta_4^p-\frac{1}{2}\beta_{3,EW}^p-\frac{1}{2}\beta_{4,EW}^p] \nonumber \\
&+B_{ K^* K_0^*}[b_4^p-\frac{1}{2}b_{4,EW}^p]\,,  \\
{\cal A}_{\bar B_s^0 \to  K_0^{*-} K^{*+}}=&A_{K^{*} K_0^{*}}[\delta_{pu}\alpha_1+\alpha_4^p+\alpha_{4,EW}^p+\beta_3^p+\beta_4^p-\frac{1}{2}\beta_{3,EW}^p-\frac{1}{2}\beta_{4,EW}^p]\nonumber\\
&+B_{ K_0^{*} K^{*}}[\delta_{pu}b_1+b_4^p+b_{4,EW}^p] \,,   \\
{\cal A}_{\bar B_s^0 \to  K_0^{*+} K^{*-}}=&+A_{K_0^* K^*}[\delta_{pu}\alpha_1+\alpha_4^p+\alpha_{4,EW}^p+\beta_3^p+\beta_4^p-\frac{1}{2}\beta_{3,EW}^p-\frac{1}{2}\beta_{4,EW}^p]\nonumber \\
&+B_{ K^* K_0^{*}}[\delta_{pu}b_1+b_4^p+b_{4,EW}^p]\,,   \\
{\cal A}_{\bar B_s \to K_0^{*+} \rho^-}=&A_{ K_0^*\rho}[\delta_{pu}\alpha_1+\alpha_4^p+\alpha_{4,EW}^p+\beta_3^p-\frac12\beta_{3,EW}^p]\,,\\
{\cal A}_{\bar B_s \to  K_0^{*0} \rho^0}=&\frac{1}{\sqrt2}A_{ K_0^*\rho}[\delta_{pu}\alpha_2 -\alpha_4^p+\frac32\alpha_{3,EW}^p+\frac12\alpha_{4,EW}^p-\beta_3^p+\frac12\beta_{3,EW}^p]\,,\\
 {\cal A}_{ \bar B_s \to  K_0^{*0} \omega}=& \frac{1}{\sqrt2} A_{K_0^*\omega}[\delta_{pu}\alpha_2+2\alpha_3^p+\alpha_4^p+\frac12\alpha_{3,EW}^p
-\frac12\alpha_{4,EW}^p+\beta_3^p-\frac12\beta_{3,EW}^p]\,,\\
{\cal A}_{ \bar B_s \to  K_0^{*0} \phi}=& A_{\phi K_0^*}[\alpha_4^p-\frac12\alpha_{4,EW}^p+\beta_3^p-\frac12\beta_{3,EW}^p]
+ A_{K_0^*\phi}[\alpha_3^p-\frac12\alpha_{3,EW}^p]\,.
\end{align}

\end{document}